\documentclass[12pt,a4paper]{article}
\usepackage[margin=1in]{geometry}
\usepackage[sort&compress,square,numbers]{natbib}
\usepackage{lipsum,times,graphicx}
\usepackage[labelfont=bf]{caption}
\usepackage{authblk}
\usepackage{lineno}
\usepackage{amsmath,amssymb,amsfonts}
\usepackage{amsthm}
\usepackage{mathrsfs}
\usepackage{algorithm}
\usepackage{algorithmicx}
\usepackage[version=4]{mhchem}
\usepackage{multirow}
\usepackage{booktabs}
\usepackage{soul}
\usepackage{color, xcolor}
\usepackage[colorlinks,
		linkcolor=blue,
		anchorcolor=blue,
		citecolor=blue,
		urlcolor=blue]{hyperref}

\usepackage[normalem]{ulem}

\usepackage{caption}
\usepackage{caption}
\title{
\textbf{On the Covalent Fields of Molecule–Surface Interactions}
}

\author[1,2*]{Edvin Fako}
\author[1,2*]{Philippe Schwaller}
\affil[1]{Laboratory of Artificial Chemical Intelligence (LIAC), 
  Institute of Chemical Sciences and Engineering, 
  Ecole Polytechnique F\'{e}d\'{e}rale de Lausanne (EPFL), 
  Lausanne, Switzerland}
\affil[2]{National Centre of Competence in Research (NCCR) Catalysis, 
  Ecole Polytechnique F\'{e}d\'{e}rale de Lausanne (EPFL), 
  Lausanne, Switzerland}
\affil[*]{Corresponding authors. Email: edvinfako@gmail.com, philippe.schwaller@epfl.ch}

\date{} 

\begin{document}
\maketitle

\subsection*{Abstract}

The ambiguity of the active site, the empirical status of Brønsted–Evans–Polanyi relations, and the unpredictability of linear scaling relation breakdown are three symptoms of a single representational choice: treating chemical affinity as an attribute of discrete geometric sites. Here we show that all three are resolved when chemical affinity is represented as a continuous property of the interface: the covalent field. We present a framework, Covalent Field Theory (CFT), in which active sites emerge as regions where the field sustains a bias toward bond formation beyond the thermal threshold, removing the need for geometric classification. Linear scaling relations are correlation structure in the field across probe families; their breakdown is a topological bifurcation with a precise geometric signature. Brønsted–Evans–Polanyi correlations arise from the covalent field decomposition, providing a theoretical basis for what has previously been treated as an empirical regularity, demonstrated across ~120,000 candidate pathways. Applied to a high-entropy alloy nanoparticle and a partially reduced high-entropy oxide, CFT maps these properties onto surfaces of arbitrary compositional and structural complexity.

\textbf{Keywords:} Heterogeneous catalysis, Covalent field theory, Linear scaling relationships, Surface representation, Machine-learned potentials, Catalyst design
\clearpage


\subsection*{Introduction} \label{introduction}

The active site, a discrete surface location held responsible for chemical transformation, has served as the foundational concept of heterogeneous catalysis for over a century \cite{Sabatier1911, Taylor1925, Langmuir1922, mc1913}. The term now appears in over 28,000 publications annually \cite{webofscience}, yet this ubiquity masks a profound conceptual problem: despite a century of use, no rigorous and universally applicable definition exists \cite{Vogt_2022}. As our capacity to characterize and simulate catalysts grows in scale and scope, the fragmentation of knowledge into highly specific, non-transferable case studies has become the primary bottleneck to rational materials design.

The modern era of computational catalysis is a triumph of the static-site paradigm. The quantitative power of Density Functional Theory (DFT) \cite{Hohenberg1964, Kohn1965, Payne1992} enabled the discovery of linear scaling relationships (LSR) and Brønsted–Evans–Polanyi (BEP) relations \cite{AbildPedersen2007, Busch2015, Bligaard2004}, transforming heterogeneous catalysis from qualitative observation to quantitative prediction \cite{Hammer2000, Medford2015, GarcaMuelas2019}. Yet these advances rest on the assumption that the catalyst surface is an (quasi) immutable, static entity. A concession that was computationally necessary but that has since become structurally embedded in how data is generated, interpreted and modeled, despite being at odds with decades of experimental evidence for dynamic surface reconstructions under reaction conditions \cite{Ertl1967-qc, Somorjai1989, Somorjai1988}.

Despite a wealth of experimental insight, and as a direct consequence of the static-site paradigm, the field faces a self-reinforcing problem. The current computational toolkit provides no way of measuring bond-forming capacity directly: reactivity is either inferred from observation (molecular dynamics, metadynamics) or reconstructed event-by-event by explicitly connecting local minima through transition-state theory \cite{Truhlar1996}. Despite sophisticated methods for evaluating ensemble energetics (via quantum chemistry \cite{Hohenberg1964, Kohn1965} or machine learning \cite{Bartk2018, mace2022}) and for propagating the system through phase space, the computational study of surface reactivity remains an observational science. Data-driven methods and machine learning promise to accelerate discovery, but largely rely on datasets generated within the same paradigm \cite{ocp_dataset, oc22_dataset, odac23_dataset}, biasing models toward idealized systems and limiting their ability to reason and discover \cite{Bozal-Ginesta2025-lu, Xin2025-roadmap}. These are not independent shortcomings; they share a common origin in the representational language the field has inherited. Heterogeneous catalysis shares the same atomistic representation with alloy, battery, and fluid research, yet the very definition of a chemical reaction involves a change in the chemical identity of a molecular species. Within the atomistic representation, the static-site idea is not an added assumption but a necessity: chemical species are identified as arrangements of atoms corresponding to local minima of the potential energy surface. As a direct consequence, simple molecules, high-symmetry surfaces, and high-symmetry sites are over-represented not because they are the most relevant, but because they are the easiest to construct and label reliably.

The rapid development of machine-learned interatomic potentials \cite{Bartk2018, mace2022, Chen2022, Behler2024} has made simulation of structurally and compositionally complex surfaces increasingly tractable \cite{Han2025, Han2023-zm}, shifting the primary bottleneck in computational catalyst discovery from energy evaluation to representation, and crucially, interpretation. Existing spatial descriptions of surface reactivity do not fill this gap. Generalized coordination numbers provide geometry-based affinity proxies without chemical specificity \cite{CalleVallejo2015}; local d-band center \cite{Hammer1995} maps offer spatially resolved electronic descriptors \cite{Norskov1980,Norskov1982} but remain properties of the reference surface alone, with probe identity entering only implicitly. Dense adsorption energy grids achieve chemical specificity but require ionic relaxation at every point, making them computationally prohibitive and inherently incompatible with dynamic or metastable surfaces \cite{Margraf2023, Fako2026aads}. What is needed, and what has not previously existed, is a description that is simultaneously probe-specific, energetically interpretable, applicable to surfaces of arbitrary complexity, and evaluable at non-equilibrium configurations without requiring relaxation to local minima or statistical sampling of rare reactive events.

In this contribution, we show that assuming surfaces possess an intrinsic capacity for hosting and facilitating reactive events that varies continuously across the interface, leads to a unified picture of surface reactivity. We introduce Covalent Field Theory to quantify this capacity, defined at arbitrary configurations, and without requiring relaxation of the surface or prior identification of binding sites. The ability to evaluate the covalent field independently of the instantaneous configuration of the surface unlocks the ability to monitor how reactivity evolves as the surface reconstructs over time. The covalent field is defined at any point within the system, but can be evaluated on an orientable two-dimensional manifold enclosing the interface, sampled systematically without relaxation or prior site identification.

The practical need for exactly this description first surfaced in an experimental collaboration on ligand-modulated copper electrocatalysts for CO$_2$ reduction \cite{Leemans2026} (Supplementary Note~1), where treating affinity as a continuous field resolved a question that the discrete-site framework had 
no natural language for. That result, obtained before the framework was formally axiomatized, was the first indication that the representational choice carries genuine explanatory power.

For trivial (atomic) probes the covalent field is a scalar field on this manifold; for probes of increasing internal complexity it extends naturally to a systematic hierarchy of richer representations, defined formally in the theory section. Reactive events can also be represented with covalent fields. In the initial and final states of a surface reaction, the total energy is well described by the individual covalent fields of the participating probes. As probes approach each other along a reaction coordinate, however, their fields begin to couple and the pairwise description breaks down. The deviation from additivity (the higher-body-order residual) reaches a maximum at a configuration that places the greatest simultaneous covalent demand on a single surface site. This configuration, the point of maximal deviation, can itself be represented as a probe with its own covalent field: a spatially resolved map of the surface capacity to facilitate bond formation and breaking.

From the structure of the covalent field, catalytically relevant observables emerge as derived quantities rather than geometric assumptions. Linear scaling relations arise as correlations between probe-specific fields across the same surface, a manifold-level property that holds even when every site is chemically distinct. Even the surface capacity to facilitate bond formation and breaking has its own covalent field that couples the fields of reactants and products, and from which unit-slope Brønsted–Evans–Polanyi relations emerge as an exact algebraic identity. This allows for separation of two contributions heavily confounded in conventional analysis, an intrinsic site-dependent covalent barrier and a configurational term from the initial state.

The demonstrations that follow apply CFT to surfaces of genuine chemical complexity, a high entropy alloy nanoparticle and a partially reduced high entropy oxide, where conventional site enumeration is intractable, making the field representation the natural language of analysis. Across these systems, linear scaling relations and their deviations are resolved spatially, competing adsorption requirements are mapped simultaneously onto surface geometry, and the intrinsic site dependent barrier is separated from configurational initial state contributions across ~120,000 candidate reaction pathways. In this language, the active site finds a definition it has lacked for a century: a region where the covalent field sustains a bias toward bond formation large enough to be catalytically decisive.

\subsection*{Defining the Covalent Field}

\begin{figure}[t!]
\centering
\includegraphics[width=1.0\linewidth]{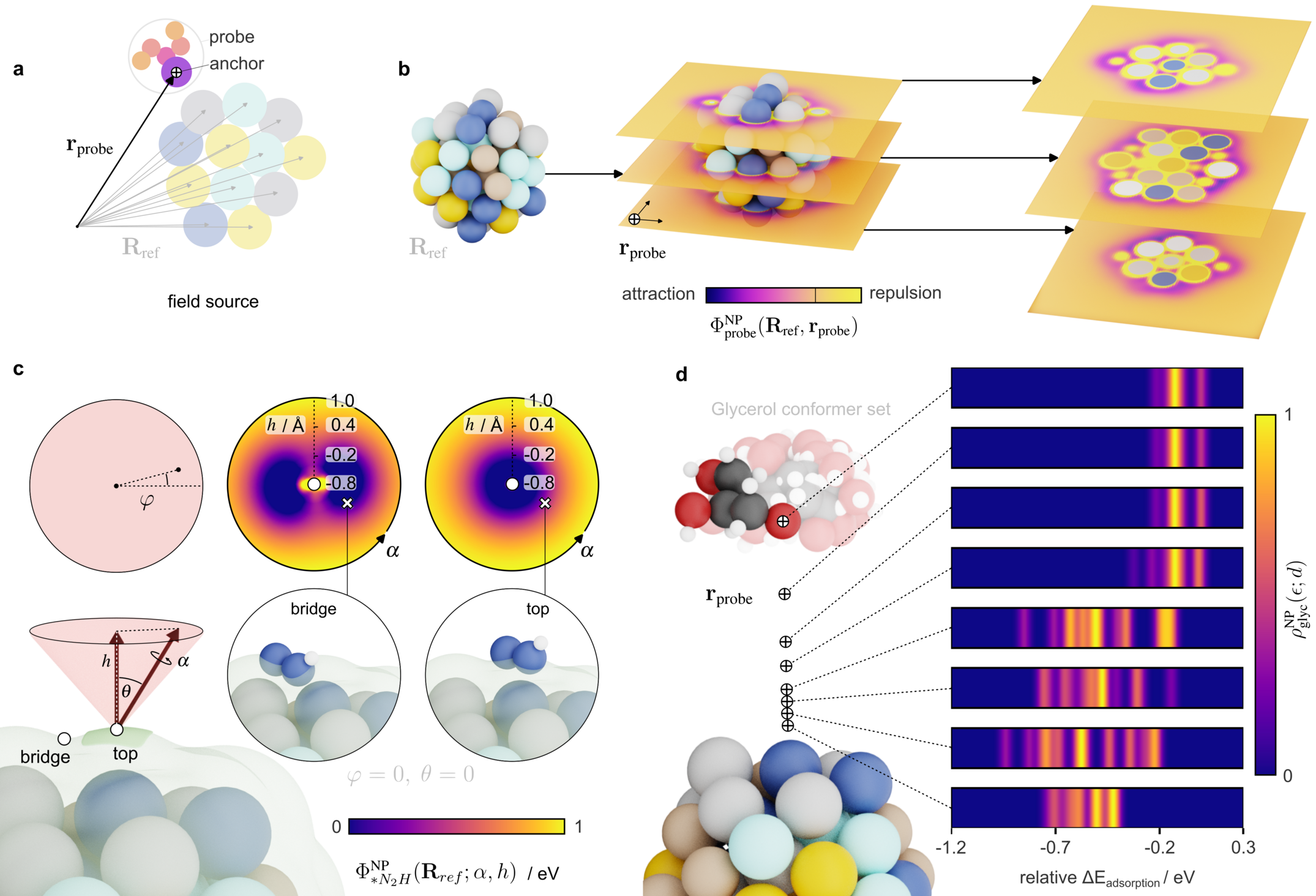}
\caption{\textbf{Covalent field decomposition of an interface interaction}
\textbf{a}, CFT partitions any surface-adsorbate system into a 
field-generating reference $\mathcal{R}$ and a molecular probe 
$\mathcal{G}$; their interaction energy, evaluated as the probe 
moves through space, defines the covalent field 
$\Phi^{\mathcal{R}}_{\mathcal{G}}$.
\textbf{b}, Covalent field $\Phi^{\mathcal{R}}_{*\text{C}}$ 
mapped around the PtPdAuAgCu nanoparticle in three planar 
slices; yellow regions are repulsive, blue regions are 
strongly attractive, illustrating the local character of 
covalent interactions.
\textbf{c}, Anisotropic field for $*\text{NNH}$ as a function 
of distance from the surface $h$ and rotation angle $\alpha$ 
about the surface normal ($\varphi=\theta=0$). The field is 
nearly isotropic at a top site but varies by $\sim$0.5~eV 
at a bridge site depending on whether the N--N axis is 
parallel or orthogonal to the metal--metal bond, illustrating 
the orientational sensitivity captured by the anisotropic 
field extension.
\textbf{d}, Spectral field for glycerol 
$\rho^{\mathcal{R}}_{\mathcal{G}}(\epsilon; d)$ represented 
as a kernel density of relaxed conformer energies at each 
probe-surface distance. Far from the surface the distribution 
reflects gas-phase conformational preferences; near the 
surface, probe-surface interactions narrow the accessible 
conformational space, selecting specific interaction modes.}
\label{fig:main}
\end{figure}

Every computational study of surface reactivity implicitly performs the same decomposition: a surface is prepared, a molecule is placed near it, and an interaction energy is computed. Covalent Field Theory makes this decomposition explicit and systematic. Any surface-adsorbate system is partitioned into a \textit{reference structure} $\mathcal{R}$ (a particle, slab, or any other material interface) and a chemical \textit{probe} $\mathcal{G}$ (an atom, molecule, or molecular fragment) defined by its molecular graph (\textbf{Figure~\ref{fig:main}a}). The interaction energy between probe and reference, evaluated as the probe moves through the space surrounding the interface, defines a continuous map of surface chemical affinity: the \textit{covalent field} $\Phi^{\mathcal{R}}_{\mathcal{G}}(\mathbf{r})$. This field is the primary object of the theory; all catalytically relevant quantities are derived from its structure rather than imposed upon it (Postulate~III, Methods).

The probe's position in space is represented by a single point, the \textit{anchor}: the atom (or geometric midpoint of atoms) that forms the primary covalent bond to the surface. This reduction is not a simplification but a consequence of the nature of covalent bonding, which is both local and directional: all properties of the probe, its orientation, conformation, and internal geometry, project onto this single point without loss of the dominant interaction (Postulate~I, Methods). The probe is oriented at each anchor position facing away from the reference, following the covalent interaction vector of the surface bond~\cite{Fako2026aads} (see Methods). The locality of covalent interactions has a second structural consequence: interaction energies decay rapidly beyond bonding distances, confining the chemically relevant domain of the field to a thin shell enclosing the interface. This shell is represented as an orientable two-dimensional manifold $\mathcal{M}$, constructed by fitting a mesh to the reference geometry. Evaluating the covalent field at each manifold vertex yields a spatial map of surface affinity that is directly interpretable in energetic terms: each vertex corresponds to the interaction energy of a specific probe-reference configuration. Because the field is a smooth, continuous function on this manifold, its critical points, 
the binding minima, migration saddles, and repulsive maxima, are not independent: their counts are linked through a global topological identity that constrains the active site structure of the entire surface simultaneously (Methods). Active sites cannot be created or destroyed in isolation; a new binding minimum must be accompanied by a new saddle connection, making the active site count a global property of the field rather than a local geometric one.
 
 
The simplest realization of the covalent field is the \textit{frozen probe} approximation, in which a single representative conformation defines the probe identity and the field becomes a scalar function of anchor position alone. This is the zeroth order term of a systematic hierarchy that extends naturally to a fiber bundle~\cite{Fatibene2003} over orientation and conformational space. When orientational effects are significant, retaining the dependence on rotation $\alpha$ about the surface normal at each manifold vertex yields the \textit{anisotropic field} $\Phi^{\mathcal{R}}_{\mathcal{G}}(\mathbf{r}; \alpha)$, capturing the directional character of probe-surface coupling that the scalar field does not. When the probe has rich conformational freedom, the distribution of interaction energies over a relaxed conformer ensemble at each anchor position defines the \textit{spectral field} $\rho^{\mathcal{R}}_{\mathcal{G}}(\epsilon; \mathbf{r})$. Both extensions are illustrated in \textbf{Figure~\ref{fig:main}c,d} and defined formally in Methods.
 
When multiple probes occupy the surface simultaneously and their separation is large relative to the range of covalent interactions, the total energy is well approximated by the sum of their individual fields. This additivity condition defines the regime of validity of the field description (Postulate~IV, Methods). Along a reaction coordinate, however, the molecular graphs of the participating probes change and the probes interact directly: the pairwise description breaks down and the total energy deviates from the sum of individual fields. The \textit{higher-body-order residual} $\Delta E_{\text{HBO}}$ 
is the exact remainder of this deviation by construction, not an approximation, and its magnitude along the reaction coordinate quantifies the degree of inter-probe coupling at each configuration. 

When a bond forms, the molecular graph transitions between well-defined CFT probes, of the reactants on one, and products on the other side. The probe at this transition, whose anchor is the forming bond, has its own covalent field that spatially represents how the surface couples the fields of reactants and products. Evaluated across the manifold, $\Delta E_{\text{HBO}}^{\mathcal{M}}$ of this probe identifies regions of minimum higher-body-order repulsion, the intrinsic barrier field, from which unit-slope Brønsted–Evans–Polanyi relations follow as an exact algebraic identity, derived formally in Methods. In deriving this identity, the assumption is made that multiple initial/final state pairs traverse through the same minimal energy pathway. However, the covalent fields of the reactants, PMD probe and products contain information about all possible pathways.

The covalent field is oracle-agnostic: the field topology, scaling relation structure, and $\Delta E_{\text{HBO}}$ decomposition are independent of the choice of energy source provided the oracle preserves the ordering and curvature of interaction energies across the manifold. MACE-MH1-r2scan~\cite{Batatia2025} is used throughout as a computationally efficient and broadly validated energy source.

\section*{Results and Discussion}

\subsection*{Surface manifolds and spatially-resolved energetics}

\begin{figure}[b!]
\centering
\includegraphics[width=1.0\linewidth]{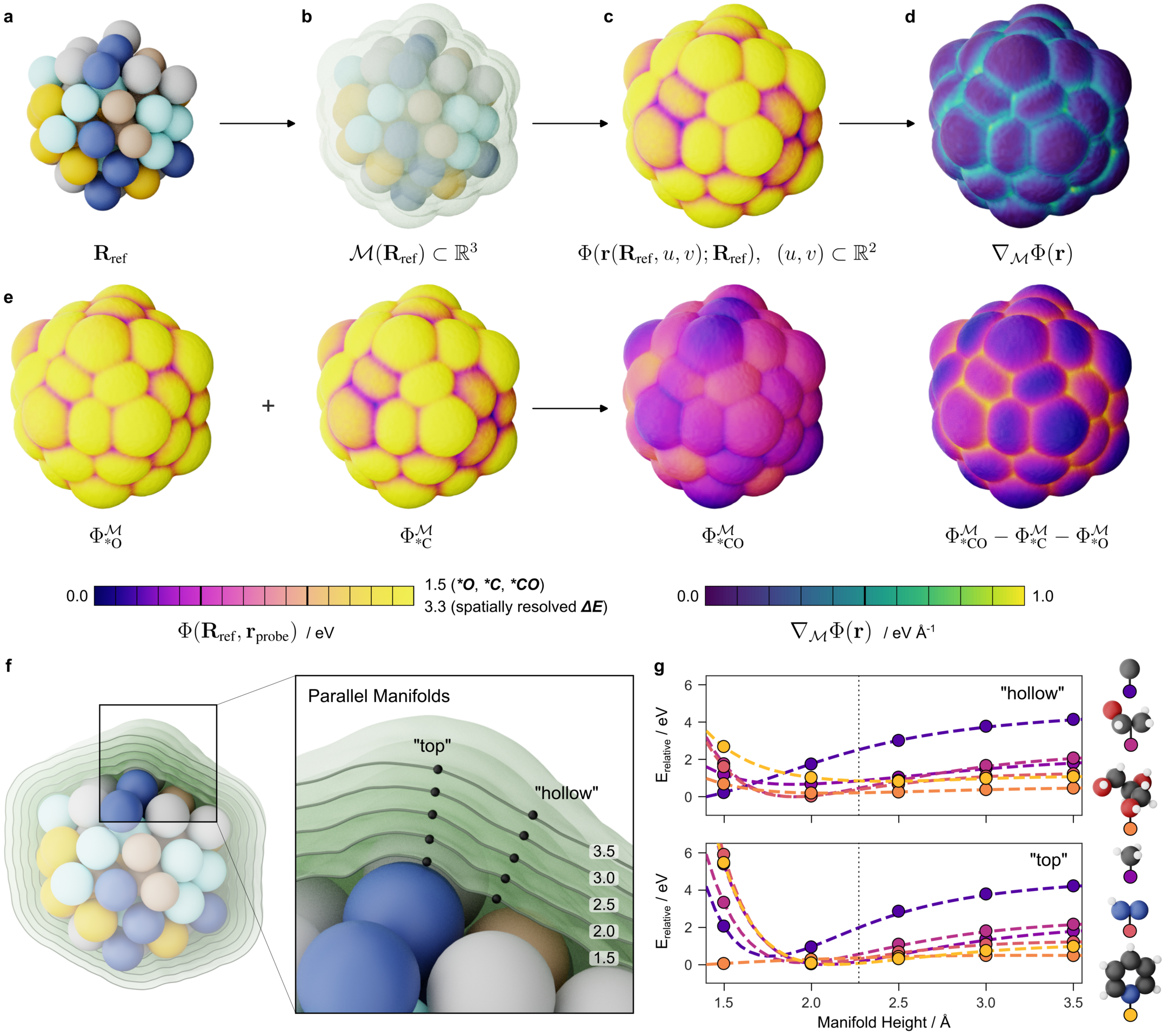}
\caption{\textbf{Covalent field manifold.}
\textbf{a}, PtPdAuAgCu high-entropy alloy nanoparticle 
containing 55 atoms.
\textbf{b}, Orientable quad mesh $\mathcal{M}$ fitted to 
the particle surface; preserved topology enables surface 
normal computation at each vertex.
\textbf{c}, Covalent field of carbon 
$\Phi^{\mathcal{M}_{\text{NP}}}_{*\text{C}}$ mapped onto 
the manifold as a function of the two-dimensional surface 
coordinates, and its gradient map (\textbf{d}); each vertex 
corresponds to the interaction energy of a specific 
probe-reference configuration.
\textbf{e}, Individual covalent fields for oxygen 
($\Phi^{\mathcal{M}_{\text{NP}}}_{*\text{O}}$), carbon 
($\Phi^{\mathcal{M}_{\text{NP}}}_{*\text{C}}$), and carbon 
monoxide ($\Phi^{\mathcal{M}_{\text{NP}}}_{*\text{CO}}$), 
and the manifold-resolved reaction energy 
$\Delta E^{\mathcal{M}_{\text{NP}}}_{\text{react.}}$, 
illustrating that additive thermodynamic quantities are 
directly defined on the manifold.
\textbf{f}, Manifolds constructed at distances spanning 
1.5 to 3.5~\AA\ from the reference particle.
\textbf{g}, Covalent field profiles as a function of 
probe-surface distance for representative top and hollow 
sites across probes of increasing complexity (atomic 
carbon, pyridine, glycerol); profiles follow Morse-like 
shapes with probe-specific minima, and ordering is 
preserved above approximately 2.3~\AA, confirming that 
the static manifold captures field topology without 
reproducing exact adsorption minima.}
\label{fig:manifold}
\end{figure}

Mapping the $*\text{C}$ covalent field across the manifold $\mathcal{M}$ of the HEA nanoparticle yields a continuous, spatially resolved affinity map $\Phi^{\mathcal{M}}_{*\text{C}}$ (\textbf{Figure~\ref{fig:manifold}c}) with well-defined gradients (\textbf{Figure~\ref{fig:manifold}d}). Because each vertex corresponds to the interaction energy of a specific probe-reference configuration, additive quantities are defined on the manifold without further approximation: the reaction energy $\Delta E_{\text{react.}}^{\mathcal{M}} = \Phi^{\mathcal{M}}_{*\text{CO}} - \Phi^{\mathcal{M}}_{*\text{C}}
- \Phi^{\mathcal{M}}_{*\text{O}}$ maps point-by-point thermodynamics directly onto the surface (\textbf{Figure~\ref{fig:manifold}e}). This description of surface resolved reaction energy ignores all geometric considerations, a point that will be revisited in detail.
 
The covalent field is a smooth function on a compact oriented manifold; the Morse inequalities \cite{Forman2002} then require that the counts of binding minima ($c_0$), migration saddles ($c_1$), and repulsive maxima ($c_2$) satisfy $c_0 - c_1 + c_2 = 2$ for a spherical nanoparticle: a global 
constraint (Methods) that makes active site counts a topological feature of the probe-surface interaction rather than a consequence of local atomic arrangement. 

The manifold is evaluated at a fixed distance from the reference particle. Fitting manifolds across 1.5 to 3.5~\AA\ and evaluating the field for probes of increasing complexity (Figure~\ref{fig:manifold}f) shows that energy profiles follow Morse-like shapes with probe-specific minima (predictably given the smoothness of the governing MLIP). Above approximately 2.3~\AA\ the profiles become uniform and well-separated, confirming that field ordering and critical point structure are preserved across the sampled range. The smooth variation in the surface-normal direction further establishes that the full distance-parameterized family is continuous in three dimensions, providing a formal basis for the distance-independence of the topological structure observed here.

\subsection*{Linear scaling relations across the manifold}

\begin{figure}[t!]
\centering
\includegraphics[width=1.0\linewidth]{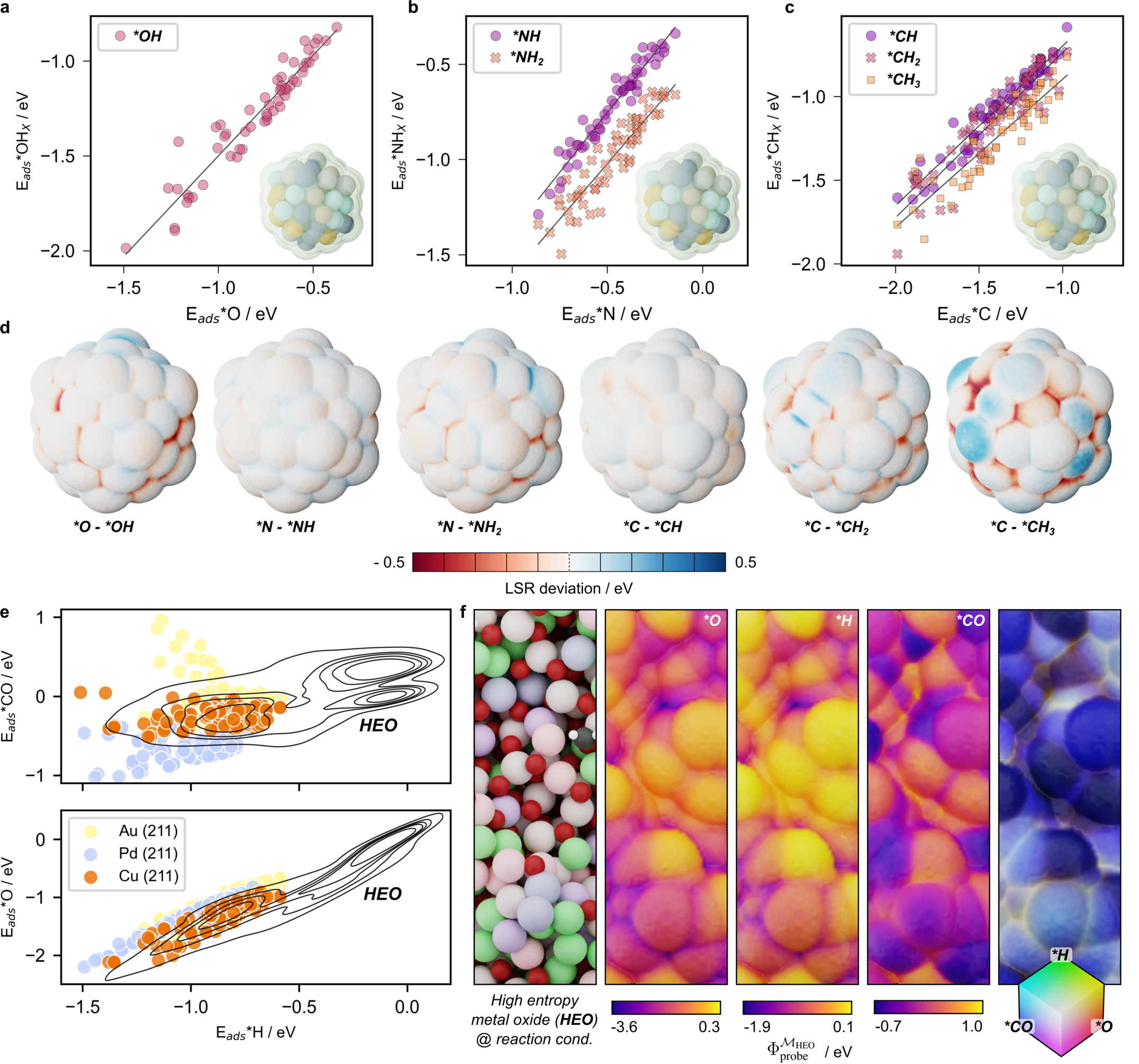}
\caption{\textbf{Linear scaling relations across the manifold.}
\textbf{a--c}, Covalent field correlations for probe families 
$*\text{O}$/$*\text{OH}$, $*\text{N}$/$*\text{NH}$/$*\text{NH}_2$, 
and $*\text{C}$/$*\text{CH}$/$*\text{CH}_2$/$*\text{CH}_3$ across 
the HEA nanoparticle manifold. Linear trends persist even though 
every surface site is chemically distinct, establishing LSRs as a 
manifold-level property rather than a property of equivalent sites.
\textbf{d}, Spatial distribution of LSR deviations projected onto 
the manifold; gray regions satisfy the scaling relation, blue (red) 
regions bind more weakly (strongly) than predicted.
\textbf{e}, Affinity distributions for $*\text{CO}$, $*\text{O}$, 
and $*\text{H}$ on the partially reduced HEO surface compared 
against Cu, Au, and Pd(211) references (black); the HEO 
distribution spans both oxidized and Cu-like affinity regimes.
\textbf{f}, Multi-channel affinity descriptor mapping 
$\Phi^{\mathcal{M}_{\text{HEO}}}_{*\text{O}}$ (red), 
$\Phi^{\mathcal{M}_{\text{HEO}}}_{*\text{H}}$ (green), and 
$\Phi^{\mathcal{M}_{\text{HEO}}}_{*\text{CO}}$ (blue) 
simultaneously onto the surface. White regions bind all three 
species strongly and correspond to partially reduced motifs 
most relevant for CO$_2$ reduction activity.}
\label{fig:lsr_manifold}
\end{figure}

Linear scaling relations arise in conventional analysis from comparisons of adsorption energies at equivalent sites across different materials: a procedure that becomes ambiguous on chemically complex surfaces where no two sites are alike. In the field representation this ambiguity dissolves: exchanging the 
molecular graph of the probe while fixing the anchor position and surface-normal orientation enforces geometric equivalence automatically, without prior site classification. Applied to three probe families ($*\text{O}$/$*\text{OH}$, $*\text{N}$/$*\text{NH}$/$*\text{NH}_2$, and 
$*\text{C}$/$*\text{CH}$/$*\text{CH}_2$/$*\text{CH}_3$) across the HEA nanoparticle, the manifold distributions follow linear trends (\textbf{Figure~\ref{fig:lsr_manifold}a--c}) even though every surface site is chemically distinct. Linear scaling relations are not a property of equivalent sites across materials, they are a manifold-level property of a single complex surface. Deviations from the linear trend are computed per vertex and projected back onto the surface (\textbf{Figure~\ref{fig:lsr_manifold}d}), spatially localizing the regions of outlier catalytic behavior that scaling-plot approaches can detect but never locate.

Balancing competing adsorption requirements across sequential reaction steps is the central challenge of multi-functional catalyst design. For CO$_2$ reduction, optimal activity requires simultaneous balance of $*\text{CX}$, $*\text{H}$, and $*\text{OX}$ affinities, a balance achieved by Cu but not Pd or Au~\cite{Bagger2017}. High-entropy oxides under reaction conditions partially reduce and reconstruct, generating surface motifs of precisely this complexity~\cite{Han2022-xf}. Comparing affinity distributions across equivalent manifolds for Cu, Au, and Pd(211) against the HEO surface (\textbf{Figure~\ref{fig:lsr_manifold}e}) shows that the expected reactivity hierarchy is preserved across the manifolds of the pristine reference metals, 
while the HEO distribution reveals spatially distinct regions of both oxidized and Cu-like affinity toward $*\text{CO}$, $*\text{O}$, and $*\text{H}$.

Mapping all three fields simultaneously onto the surface as a multi-channel descriptor ($*\text{O}$, $*\text{H}$, $*\text{CO}$ assigned to red, green, and blue channels respectively, \textbf{Figure~\ref{fig:lsr_manifold}f}) reveals that $*\text{H}$ and $*\text{O}$ affinities are strongly correlated across the surface while $*\text{CO}$ affinity varies independently. Sites binding all three species strongly appear white and correspond to (partially) reduced surface bridge and hollow motifs most relevant for CO$_2$ reduction activity. The physical origin is directly readable from the map: oxidized surface domains suppress $*\text{H}$ affinity while leaving $*\text{CO}$ affinity largely unaffected, a spatial decoupling that correlation-plot approaches average away and that the field representation makes immediately visible.

\subsection*{Decomposing a surface reaction into covalent field contributions}

\begin{figure}[ht!]
\centering
\includegraphics[width=1.0\linewidth]{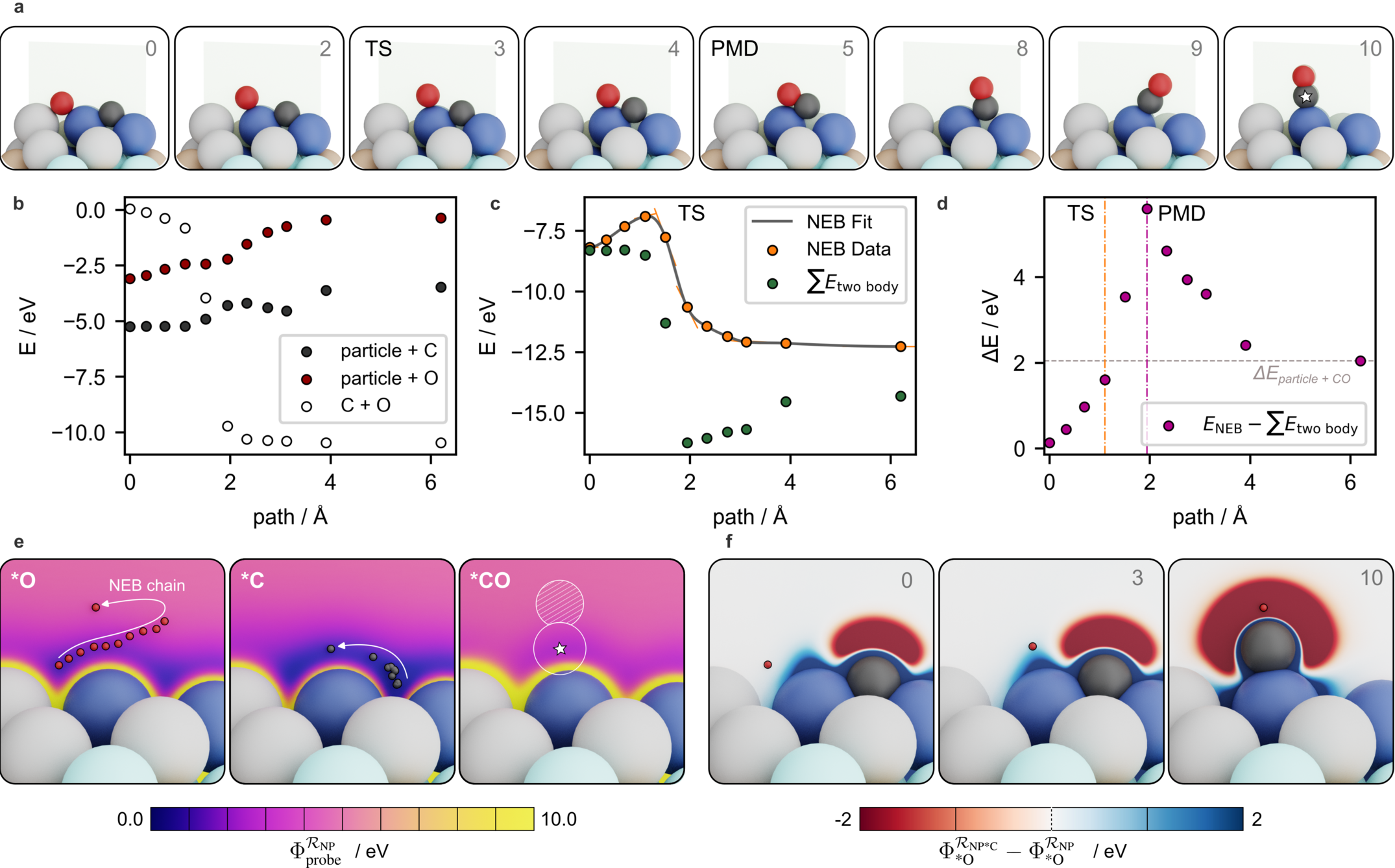}
\caption{\textbf{Covalent field decomposition of the 
$*\text{C} + *\text{O} \rightarrow *\text{CO}$ reaction pathway.}
\textbf{a}, Minimum-energy pathway computed by the nudged 
elastic band method; images are colored by total energy 
relative to the initial state.
\textbf{b}, Pairwise decomposition of the total energy into 
individual probe-field contributions: $\Phi^{\mathcal{P}_{\text{NP}}}_{*\text{C}}$ 
(gray), $\Phi^{\mathcal{P}_{\text{NP}}}_{*\text{O}}$ (red), and the C--O 
pair interaction $E_{\text{C-O}}$ (white). Both atoms are 
displaced from their equilibrium geometries along the pathway, 
with the energy penalty offset by C--O bond formation.
\textbf{c}, Sum of pairwise contributions (green) compared 
to the total energy; the pairwise description holds in the 
initial and final states but shows no barrier, demonstrating 
that the barrier originates entirely in higher-body-order terms.
\textbf{d}, Higher-body-order residual $\Delta E_{\text{HBO}}$ 
along the pathway; negligible at the initial state, rising as 
the probes couple, and reaching its maximum at image 5 --- 
the point of maximal deviation (PMD).
\textbf{e}, Trajectories of the individual probes through 
their respective covalent fields along the pathway; $*\text{O}$ 
moves substantially while $*\text{C}$ remains largely 
stationary, reflecting the stronger covalent field of carbon.
\textbf{f}, Covalent field of $*\text{O}$ in the presence 
of carbon, $\Phi^{\mathcal{P}_{\text{NP}*\text{C}}}_{*\text{O}}$, at each 
NEB image; the O position is indicated in red.}
\label{fig:neb}
\end{figure}

To build physical intuition for the CFT decomposition, we trace a single minimum-energy pathway for $*\text{C} + *\text{O} \rightarrow *\text{CO}$ on the HEA nanoparticle (\textbf{Figure~\ref{fig:neb}a}). The pairwise decomposition of the total energy into individual probe-field contributions (\textbf{Figure~\ref{fig:neb}b,c}, \textbf{\ref{SI_neb}}) reveals that the barrier originates entirely in higher-body-order terms: the pairwise contributions show no barrier.

The probe trajectories tell the physical story directly (\textbf{Figure~\ref{fig:neb}a}). Two distinct modes are visible: first, O traverses the surface toward C while remaining covalently bound to the surface, with C barely moving, reflecting the stronger covalent field of carbon. Once the C--O bond forms (\textbf{Figure \ref{SI_co_field}}), the mode changes: O lifts away from the surface as C shifts toward the Pd top site. The point separating these two modes is not the transition state, it is the configuration of greatest simultaneous coordination demand on the surface, the Point of Maximal Deviation (PMD).

The higher-body-order residual $\Delta E_{\text{HBO}}$ (\textbf{Figure~\ref{fig:neb}d}) confirms this directly: negligible at the initial state where C and O are well-separated, rising as the probes couple, and peaking precisely at the mode change. The transition state at image~3 appears earlier in the trajectory than intuition suggests, pulled toward the initial state by the steep energy release of C--O bond formation. Here the TS is significantly affected by the thermodynamics of the initial and final states.

\textbf{Figures~\ref{fig:neb}e} and \textbf{\ref{fig:neb}f} reveal the two competing field effects that govern any reaction trajectory taking place on the surface: O (or C) must escape its own covalent minimum on the interface, while the system resists the many-body repulsion arising from overcoordination of the surface atom (\textbf{Figure \ref{SI_co_field}}). The observed NEB pathway is a compromise between these two effects, but both are expressions of the covalent field. In principle, all pathways connecting initial and final states are possible; which one a reaction event follows is a statistical outcome of reaction dynamics. The covalent fields are the landscape that biases these events, all thermodynamic information about all such pathways is already contained within them.

Barriers of reactions proceeding without continuous surface contact reduce to the covalent field directly; for surface-mediated transformations, the PMD probe satisfying the covalent binding requirements of both reactants simultaneously maps the intrinsic barrier field across the full manifold, as demonstrated in the following section.

\subsection*{Mapping reaction barriers to the manifold}

\begin{figure}[ht!]
\centering
\includegraphics[width=1.0\linewidth]{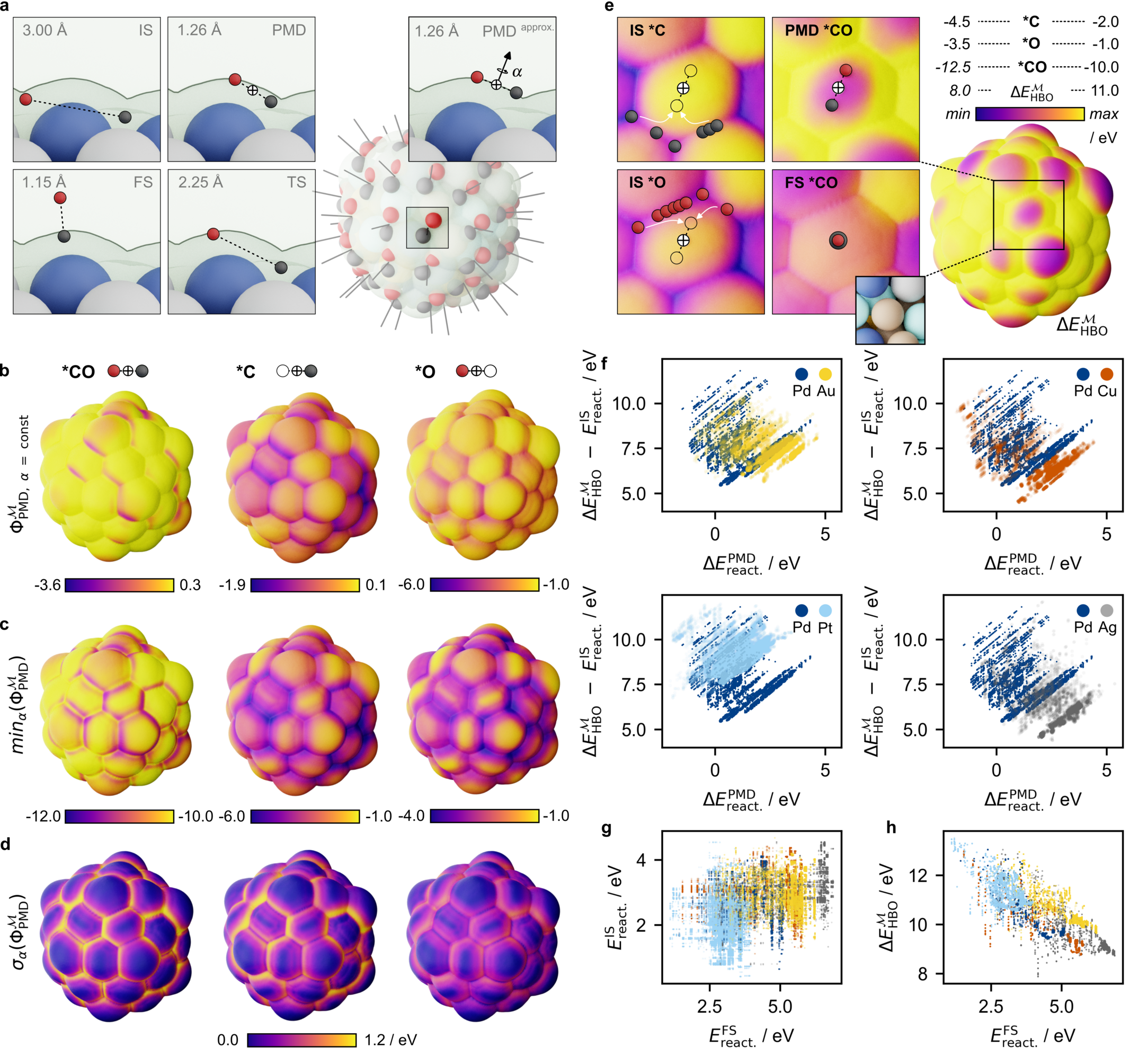}
\caption{\textbf{Brønsted--Evans--Polanyi correlations as 
a consequence of covalent field decomposition.}
\textbf{a}, PMD probe geometry identified along the $*\text{C}+*\text{O}\rightarrow*\text{CO}$ pathway; anchor 
at the geometric midpoint of the forming bond, with both 
atoms simultaneously within covalent range of the same 
surface atom.
\textbf{b}, Anisotropic covalent field of the PMD probe 
$\Phi_{\text{PMD}}^{\mathcal{M}}$ decomposed into pairwise 
contributions at fixed $\alpha$; $\Delta E_{\text{HBO}}^{\mathcal{M}}$ 
is the remainder after subtracting individual probe-field 
contributions.
\textbf{c}, Minimum of $\Phi_{\text{PMD}}^{\mathcal{M}}$ 
over $\alpha$ per manifold vertex.
\textbf{d}, Standard deviation 
$\sigma_\alpha(\Phi^{\mathcal{M}}_{\text{PMD}})$ over 
$\alpha$; bridge and three-fold sites show variations up 
to $\sim$1~eV, providing a diagnostic of frozen 
approximation reliability.
\textbf{e}, Manifold distribution of 
$\Delta E_{\text{HBO}}^{\mathcal{M}}$; purple basins 
identify candidate active sites capable of sustaining 
simultaneous dual coordination.
\textbf{f}, Thermodynamics-corrected barriers 
($\Delta E_{\text{HBO}}^{\mathcal{M}} - 
E_{\text{react.}}^{\text{IS}}$) versus reaction energy 
across $\sim$120,000 candidate pathways, colored by 
anchoring metal. Unit-slope parallel families emerge 
as an exact algebraic identity of the decomposition; 
vertical offsets count topologically distinct active 
site environments. Pd (dark blue) shown as reference.
\textbf{g}, Initial-state versus final-state energies; 
absence of correlation confirms IS configurations are 
distributed independently of the basin they access.
\textbf{h}, Final-state $*\text{CO}$ affinity versus 
$\Delta E_{\text{HBO}}^{\mathcal{M}}$; inverse correlation 
reflects geometric incompatibility between PMD dual 
coordination and the top-site preference of $*\text{CO}$.}
\label{fig:bep_manifold}
\end{figure}

The preceding section showed that for a reaction that proceeds on an interface, where neither reactant loses covalent contact to the surface, a configuration exists for which the molecular graphs of the reactants transform into the molecular graph of the product. This configuration, point of maximal deviation (PMD) meets all CFT formal requirements to be considered itself a probe: it is defined by its own molecular graph at the transition between reactant and product fields and anchored at the forming bond midpoint. This frozen probe geometry can be constructed from covalent requirements alone and compared to the PMD configuration obtained in the previous section (\textbf{Figure~\ref{fig:bep_manifold}a}). As such, its covalent field can be calculated independently from any initial or final resting state assumptions, and the intrinsic barrier ($\Delta E_{\text{HBO}}^{\mathcal{M}}$) can be computed from the PMD covalent field (\textbf{Figure~\ref{fig:bep_manifold}b}). By decomposing the total interaction into pairwise contributions at geometrically equivalent structures, we can construct the higher-body-order deviation of the PMD probe:

\begin{equation}
    \Delta E_{\text{HBO}}^{\mathcal{M}} =
    \Phi^{\mathcal{M}}_{\text{PMD}_{*\text{CO}}}
    + E_{\text{form.PMD}}^{\text{CO}}
    - \Phi^{\mathcal{M}}_{\text{PMD}_{*\text{C}}}
    -\Phi^{\mathcal{M}}_{\text{PMD}_{*\text{O}}}
    \label{eq:many_body_bep_map}
\end{equation}

Practically, this is done by rotating the PMD probe about the surface normal in $10^\circ$ increments yielding the anisotropic field $\Phi_{\text{PMD}}^{\mathcal{M}}(\alpha)$. The minimum energy across all orientations (\textbf{Figure~\ref{fig:bep_manifold}c}) and its standard deviation $\sigma_\alpha(\Phi^{\mathcal{M}}_{\text{PMD}})$ (\textbf{Figure~\ref{fig:bep_manifold}d}) are recorded at each vertex; bridge and three-fold sites exhibit the highest orientational sensitivity, with energy variations reaching $\sim$1~eV. Because the PMD probe geometry is fully specified at each vertex and orientation, $\Delta E_{\text{HBO}}^{\mathcal{M}}$ follows directly from equation~\eqref{eq:many_body_bep_map}: the individual probe-field contributions $\Phi^{\mathcal{M}}_{\text{PMD}_{*\text{C}}}$ and $\Phi^{\mathcal{M}}_{\text{PMD}_{*\text{O}}}$ are evaluated at the atomic positions within the frozen PMD geometry and subtracted from $\Phi_{\text{PMD}}^{\mathcal{M}}$. This intrinsic barrier field, reveals discrete basins of dual-coordination capacity on the manifold (\textbf{Figure~\ref{fig:bep_manifold}e}). Purple regions identify where a single surface atom can simultaneously satisfy the covalent demands of both reactants (the geometric prerequisite for bond formation) and constitute the candidate active areas for this reaction class.

Identifying basins that satisfy the covalent requirement of the PMD establishes whether the surface can facilitate bond formation. However, the activation and reaction energies depend on the specific configurations of the initial and final states. For a given PMD vertex and orientation $\alpha$, the positions of the constituent C and O atoms within the frozen PMD geometry define candidate initial-state positions for $*\text{C}$ and $*\text{O}$, and a candidate final-state position for $*\text{CO}$, by identifying the nearest manifold vertices to each atomic anchor. Restricting to vertex pairs whose separation is sufficient for covalent field independence (\textbf{Figure~\ref{fig:bep_manifold}e}, equation~\eqref{eq:additivity}), the reaction energy for each PMD configuration is defined entirely from manifold distributions of the individual covalent fields:

\begin{equation}
    \Delta E_{\text{react.}}^{\text{PMD}} = 
    \Phi^{\mathcal{M}}_{*\text{CO}}(\mathbf{r}_{*\text{CO}}) 
    + E_{\text{form.}}^{\text{CO}} - 
    \Phi^{\mathcal{M}}_{*\text{C}}(\mathbf{r}_{*\text{C}})
    - \Phi^{\mathcal{M}}_{*\text{O}}(\mathbf{r}_{*\text{O}})
\end{equation}

By identifying manifold minima of $\Delta E_{\text{HBO}}^{\mathcal{M}}$ for each sampled orientation of the PMD probe and constructing initial and final states as described above, the 55-atom nanoparticle yields $\sim$120,000 candidate reaction pathways without transition-state searches. For each pathway, the geometries of the initial state, PMD, and final state are explicitly defined within the CFT framework. The covalent fields of the reactants and products are coupled through all of these pathways simultaneously; which one a reactive event follows is a statistical outcome of the dynamics on the covalent field landscape.

With both a thermodynamic and a kinetic observable defined on the manifold for every candidate pathway, the relationship between reaction energy and activation barrier can be examined directly. Plotting the thermodynamics-corrected barrier ($\Delta E_{\text{HBO}}^{\mathcal{M}} - E_{\text{react.}}^{\text{IS}}$) against the reaction energy (\textbf{Figure~\ref{fig:bep_manifold}f}) reveals a structure of parallel BEP families with unit slope. By construction, the unit slope is an exact algebraic identity of the decomposition (Methods): the PMD position is selected as minima of each $\Delta E_{\text{HBO}}^{\mathcal{M}}$ basin, and varying the initial state shifts both the apparent barrier and the reaction energy by equal amounts. The construction does not guarantee that each proposed barrier reflects the true minimum-energy pathway; what it establishes is that a surface moiety capable of funneling a distribution of initial states into a distribution of final states necessarily expresses itself as a family of parallel, unit-slope BEP lines. The number of such families is therefore a direct count of topologically distinct active site environments on the manifold. 

The absence of clear correlation between initial-state and final-state energies (\textbf{Figure~\ref{fig:bep_manifold}g}) demonstrates the independence of $\Delta E_{\text{HBO}}^{\mathcal{M}}$ from initial-state configuration that the decomposition requires. An inverse correlation between $\Delta E_{\text{HBO}}^{\mathcal{M}}$ and the final-state affinity for $*\text{CO}$ (\textbf{Figure~\ref{fig:bep_manifold}h}) shows that sites most effective at stabilizing the strained PMD geometry bind the $*\text{CO}$ product more weakly. This anti-correlation reflects a geometric incompatibility between the simultaneous dual coordination demanded by the PMD and the top-site preference of $*\text{CO}$: a spatial relationship invisible to site-averaged analysis but directly readable from the manifold representation.

\section*{Conclusions}

The ambiguity of the active site, the empirical status of Brønsted--Evans--Polanyi relations, and the unpredictability of linear scaling relation breakdown have resisted resolution for decades not because they are intrinsically complex, but because the discrete-site language makes them appear so. Each dissolves when chemical affinity is treated as a continuous field property of the interface rather than an attribute of discrete geometric sites.

In the field language, active sites are not located but defined: regions where the covalent field sustains a bias toward bond formation large enough to be catalytically decisive. Their counts are constrained by a global topological identity, a new binding minimum cannot appear without a corresponding saddle connection, making active site structure a property of the field topology rather than of local atomic arrangement. Linear scaling relations are correlation structure in the field across probe families, holding manifold-wide even when every surface site is chemically distinct, and their breakdown is a topological bifurcation with a precise geometric signature on the surface. Brønsted--Evans--Polanyi correlations with unit slope are an algebraic identity of the covalent field decomposition: the thermodynamics-corrected activation barrier $\Delta E_{\text{HBO}}^{\mathcal{M}}$ is independent of initial-state configuration by construction, and varying the initial state shifts both the apparent barrier and the reaction energy by equal amounts. The 120,000-pathway demonstration does not prove this slope, it confirms the self-consistency of the frozen probe approximation across a chemically heterogeneous surface.

These are not three separate results. They are three consequences of a single representational choice, visible simultaneously once the field is the primary object. The number of BEP lines counts the topologically distinct active site environments; their offsets measure intrinsic barrier capacity; their slopes are exact. The multi-channel affinity descriptor maps competing adsorption requirements directly onto surface geometry, identifying regions where the field simultaneously satisfies multiple thermodynamic criteria: a spatial specificity that correlation-plot approaches cannot provide.

Critically, the covalent field is not a static construct. While the demonstrations here use static surface snapshots for clarity, the field is defined at arbitrary configurations of the surface and probe, without requiring relaxation or prior site identification. A working catalyst is a time-evolving interface, and CFT provides the instrument for evaluating how its affinity landscape evolves as the surface reconstructs, activates, and responds to reaction conditions. The framework changes not only how efficiently surface reactivity can be described, but what questions can be asked: spatial active site topology, distance-resolved field continuity, and the topological signature of scaling relation breakdown are questions that the discrete-site 
language cannot even formulate.

As machine-learned interatomic potentials make structurally and compositionally complex surfaces increasingly tractable, the primary bottleneck in computational catalyst discovery shifts from energy evaluation to representation. Covalent Field Theory provides that representation --- not as a descriptor scheme or a computational workflow, but as a theory in which the defining characteristics of catalytic activity emerge from the structure 
of the field rather than from assumptions about the structure of the surface.



















\section*{Methods}

\subsection*{Mathematical framework}

\textbf{Configuration mapping.} The reference structure $\mathcal{R}$ is defined by the atomic positions and species of the surface or nanoparticle. The probe $\mathcal{G}$ is defined by its molecular graph, with a designated \textit{anchor} at the atom or geometric midpoint of the atom pair forming the primary covalent bond to the reference. Given anchor position $\mathbf{r} \in \mathbb{R}^3$ and orientation $\mathbf{Q} \in SO(3)$, the Cartesian position of probe atom $i$ is
\begin{equation}
  \mathbf{r}^i_{\text{probe}}(\mathbf{r}, \mathbf{Q}, \chi)
  = \mathbf{r} + \mathbf{Q}\,\boldsymbol{\xi}_i(\chi),
  \label{eq:config_map}
\end{equation}
where $\boldsymbol{\xi}_i(\chi)$ is the position of atom $i$ in the probe's body frame (anchor at origin) and $\chi$ parameterizes internal conformational degrees of freedom. In the default \textit{frozen-probe} approximation, $\mathbf{Q}$ is determined by the local surface normal and $\chi$ is fixed to a reference conformer generated by \texttt{AutoAdsorbate}~\cite{Fako2026aads}, making the probe configuration a deterministic function of $\mathbf{r}$ alone. The full set of probe atom positions under this mapping is denoted $\mathcal{P}_{\mathcal{G}}(\mathbf{r})$.

\textbf{Postulate I: Anchor definition.} Every probe has a designated anchor, specified in its molecular graph. For monodentate binding the anchor is the surface-binding atom; for bidentate binding it is the geometric midpoint of the two binding atoms. The position of the probe in space is defined as the position of its anchor, $\mathbf{r} \in \mathbb{R}^3$.

\textbf{Postulate II: Configuration mapping.} All configurations of the probe that preserve its molecular graph, rotations, conformational changes, local dynamics, map to the anchor position via equation~\eqref{eq:config_map}. Transformations that change the molecular graph define a distinct probe and a distinct covalent field.

\textbf{Postulate III: Covalent field.} The covalent field is the interaction energy between probe and reference as a function of anchor position:
\begin{equation}
  \Phi^{\mathcal{R}}_{\mathcal{G}}(\mathbf{r})
  = E\!\left(\mathcal{R} \cup 
  \mathcal{P}_{\mathcal{G}}(\mathbf{r})\right)
  - E(\mathcal{R}) - E_0^{\mathcal{G}},
  \label{eq:cft_field}
\end{equation}
where $E(\cdot)$ is the total potential energy evaluated by the chosen oracle and $E_0^{\mathcal{G}}$ is the probe gas-phase reference energy. Under the frozen-probe approximation this is a well-defined function of $\mathbf{r}$ alone. The field topology, its gradients, critical points, and spatial correlations, is independent of the choice of oracle provided the oracle faithfully reproduces the ordering and curvature of interaction energies.

\textbf{Postulate IV: Field validity and additivity.} The covalent field correctly describes probe--reference interactions when (i) the probe's molecular graph is well-defined, and (ii) inter-probe interactions are negligible. For $n$ isolated probes satisfying both conditions, the total energy is approximately additive:
\begin{equation}
  E_{\text{total}} \approx E(\mathcal{R})
  + \sum_{i=1}^{n} 
  \Phi^{\mathcal{R}}_{\mathcal{G}_i}(\mathbf{r}_i).
  \label{eq:additivity}
\end{equation}
When equation~\eqref{eq:additivity} breaks down, as along a reaction pathway where molecular graphs change, the higher-body-order residual
\begin{equation}
  \Delta E_{\text{HBO}}
  = E_{\text{total}} - E(\mathcal{R})
  - \sum_{i} 
  \Phi^{\mathcal{R}}_{\mathcal{G}_i}(\mathbf{r}_i)
  \label{eq:many_body}
\end{equation}
quantifies inter-probe coupling. Along a reaction coordinate $\Delta E_{\text{HBO}}$ reaches its maximum at the \textit{point of maximal deviation} (PMD): the configuration placing the greatest simultaneous coordination demand on a single surface atom. $\Delta E^{\text{PMD}}_{\text{HBO}}$ is obtained by evaluating equation~\eqref{eq:many_body} at the frozen PMD probe geometry across the manifold (see Point of Maximal Deviation probe construction).

\textbf{Extended representations.} When orientational effects are significant, the \textit{anisotropic field} retains the angular dependence on rotation $\alpha$ about the surface normal:
\begin{equation}
  \Phi^{\mathcal{R}}_{\mathcal{G}}(\mathbf{r};\alpha)
  = E\!\left(\mathcal{R} \cup
    \mathcal{P}_{\mathcal{G}}(\mathbf{r},\alpha)\right)
  - E(\mathcal{R}) - E_0^{\mathcal{G}}.
  \label{eq:aniso_field}
\end{equation}
When conformational sampling is required, the \textit{spectral field} represents the distribution of interaction energies over a relaxed conformer ensemble $\{\chi_k\}$:
\begin{equation}
  \rho^{\mathcal{R}}_{\mathcal{G}}(\epsilon;\mathbf{r})
  = \frac{1}{K}\sum_{k=1}^{K}
    \mathcal{N}\!\left(\epsilon;\,
      \Phi^{\mathcal{R}}_{\mathcal{G}}(\mathbf{r},\chi_k),\,
      \sigma^2\right),
  \label{eq:spectral_field}
\end{equation}
where $\mathcal{N}(\epsilon;\mu,\sigma^2)$ is a Gaussian kernel with 
bandwidth $\sigma$. The frozen-probe scalar field is the 
zero-temperature limit of equation~\eqref{eq:spectral_field}.

\subsection*{Computational tractability}

\begin{figure}[t!]
\centering
\includegraphics[width=90mm]{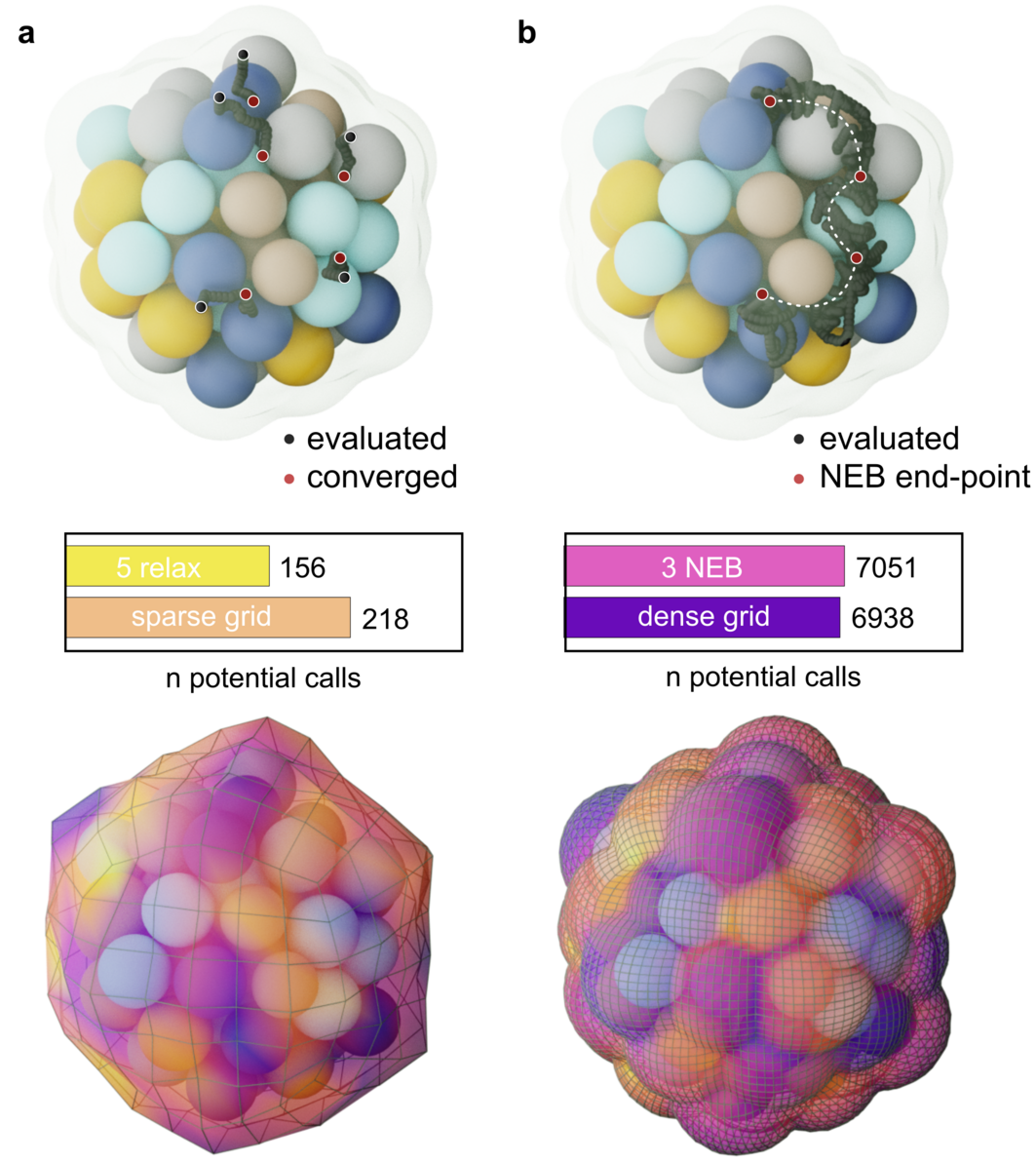}
\caption{\textbf{Computational tractability.}
\textbf{a}, Five ionic relaxations (156 calls) versus sparse manifold grid.
\textbf{b}, Three NEB calculations ($\sim$7000 calls) versus dense manifold grid. Manifold evaluation distributes effort uniformly across the surface.}
\label{fig:benchmarking}
\end{figure}

A natural question arising from the interpretability claim, where each manifold vertex corresponds to a structure whose value is its total energy (here evaluated with an MLIP), is whether this procedure is computationally tractable. Conventional workflows locate stationary points of the potential energy surface through iterative relaxations requiring tens to hundreds of potential calls. \textbf{Figure~\ref{fig:benchmarking}} compares these costs. Five ionic relaxations of $*\text{C}$ require 156 potential calls and yield five data points, whereas a sparse manifold grid (218 vertices) maps the entire surface at similar cost (\textbf{Figure~\ref{fig:benchmarking}a}). Likewise, three NEB calculations consume $\sim$7000 calls, while a dense manifold grid achieves comprehensive surface coverage at similar computational cost (\textbf{Figure~\ref{fig:benchmarking}b}). The CFT manifold approach therefore matches the cost of conventional PES exploration while distributing information uniformly across the particle surface. Rather than concentrating effort on pre-selected sites, manifold evaluation samples the surface systematically, allowing reactive regions to emerge directly from the data removing any bias in sampling.

\subsection*{Probe generation and manifold evaluation}
 
Chemical probes were constructed and placed using the \texttt{AutoAdsorbate} algorithm~\cite{Fako2026aads}, which takes a surrogate-SMILES (*SMILES) string as input, generates a reference conformer ensemble, and positions the probe on the surface by aligning the covalent interaction vector with the local surface normal at each manifold vertex. For monodentate probes the covalent interaction vector is defined by the anchor-surrogate bond; for bidentate probes it is
defined by the geometric midpoint of the two anchor atoms. Internal probe geometry was frozen at the reference conformer for all standard covalent field evaluations; conformational sampling was applied only for the spectral field calculations described in the main text.
 
The surface manifold $\mathcal{M}$ was constructed by fitting a quadrilateral mesh to the reference particle using a shrinkwrap algorithm initialized from a subdivided cube primitive. Manifolds spanning the 1.5--3.5~\AA\ distance range were generated by uniformly scaling the fitted mesh along the surface-normal direction. For each vertex, the probe anchor was placed at the vertex position, the probe was oriented according to the local normal, and the total energy of the reference-probe configuration was evaluated in a single potential call with no further relaxation. Covalent field values were obtained by subtracting the isolated reference energy $E(\mathcal{R})$ and the probe gauge energy $E_0^{\mathcal{G}}$ from the total configuration energy, following Postulate~III. Isolated reference energies were computed once per reference structure; probe gauge energies were computed for each probe species in a large simulation cell to approximate gas-phase conditions.

\begin{figure}[ht!]
\centering
\includegraphics[width=90mm]{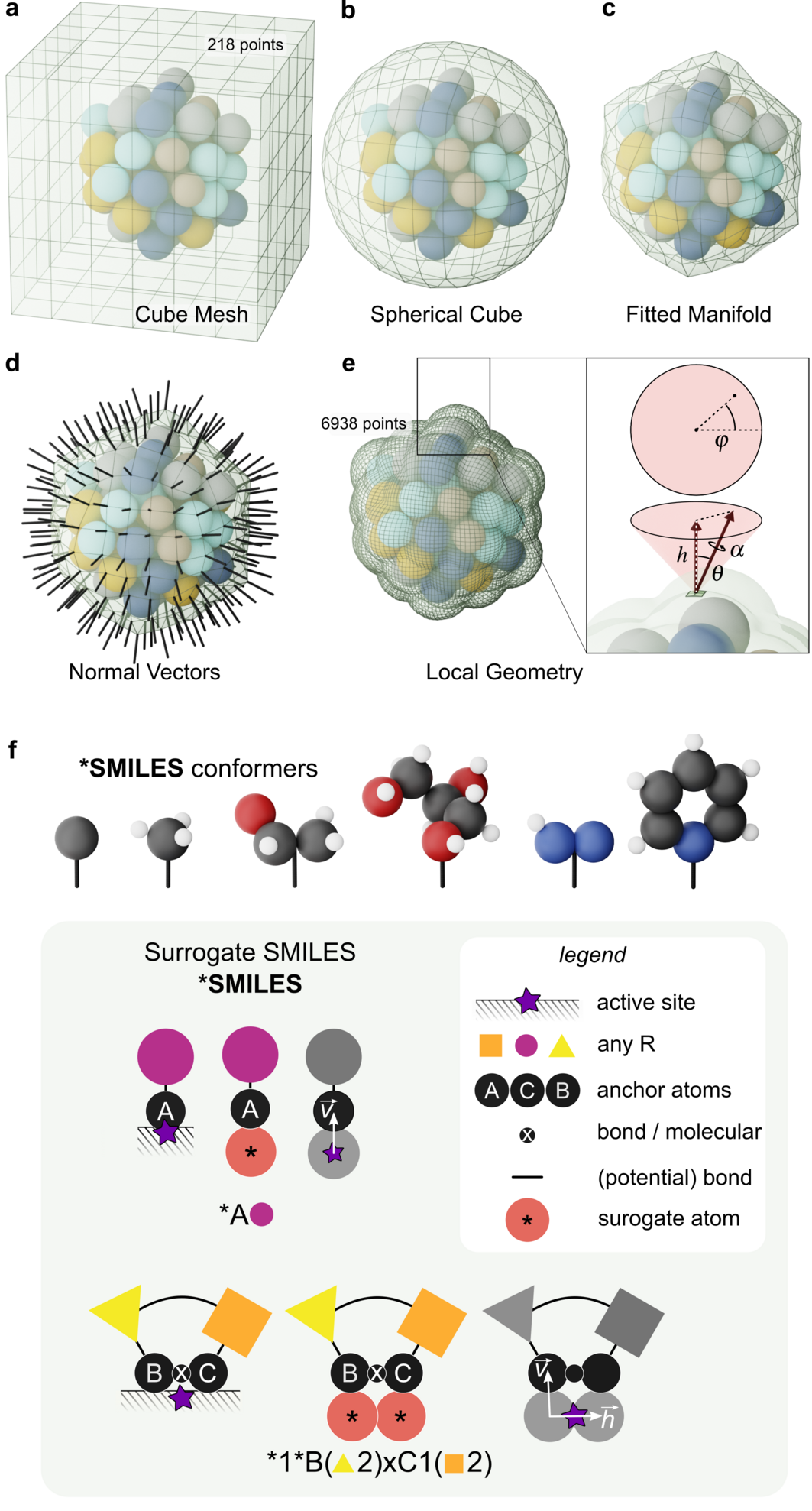}
\caption{\textbf{Manifold construction and surrogate-SMILES probe specification.}
\textbf{a}, Quadrilateral mesh fitted to the PtPdAuAgCu nanoparticle surface
using a shrinkwrap algorithm initialized from a subdivided cube primitive.
Each vertex of the cube primitive is displaced along the surface-normal direction
until contact with the particle is established, producing a mesh that conforms
to the particle geometry while preserving the original topology.
\textbf{b}, Because the mesh topology is fixed throughout the fitting procedure,
surface normal vectors are well-defined at every vertex and can be recomputed
after deformation by standard finite-difference schemes on the quad faces;
normals serve as the orientation reference for probe placement at each manifold
point.
\textbf{c}, Surrogate-SMILES (\texttt{*SMILES}) notation: a surrogate atom
(asterisk, \texttt{*}) is attached to the anchor of the probe's molecular graph,
enforcing the correct hybridization of the anchor atom and defining the covalent
interaction vector as the bond between anchor and surrogate.
\textbf{d}, Representative \texttt{*SMILES} strings and their corresponding
probe geometries after conformer generation; the surrogate atom is removed
before energy evaluation, leaving the probe oriented with its covalent
interaction vector aligned to the local surface normal.}
\label{fig:methods}
\end{figure}

\subsection*{Computational Details}

The high-entropy alloy (HEA) nanoparticle used as the primary reference structure was taken from Ref.~\citenum{ArceRamos2025}. The particle adopts a face-centred-cubic lattice with equimolar PtPdAuAgCu composition and contains 55 atoms. For the high-entropy oxide (HEO) calculations, the surface slab was taken from Ref.~\citenum{Han2025}. No additional relaxation was applied prior to CFT evaluation; the as-reported structure was used directly as the field-generating reference $\mathcal{R}$.
 
All interaction energies reported in this work were evaluated using MACE-MH1-r2scan~\cite{Batatia2025}, an equivariant message-passing neural network potential trained on a broad set of materials and molecular systems at the r\textsuperscript{2}SCAN level of density functional theory. MACE-MH1-r2scan was selected because its training distribution covers transition-metal surfaces, alloys, and oxide chemistries relevant to the systems studied here, and because its energy and force predictions have been validated against DFT reference data across the periodic table. Within the CFT framework, the potential serves as an energy oracle: the validity of the representational claims (scaling relation topology, higher-body-order residuals, and manifold affinity gradients) depends on the oracle faithfully reproducing the ordering and curvature of the interaction energy surface rather than on absolute energetic accuracy. All energy evaluations were performed using the \texttt{mace-torch} implementation interfaced through the Atomic Simulation Environment (ASE)~\cite{ase2017}.

Local minima for adsorbed intermediates were located using the Broyden-Fletcher-Goldfarb-Shanno (BFGS) optimizer as implemented in ASE, with a convergence criterion of 0.02~eV~\AA$^{-1}$ on the maximum atomic force. Minimum-energy pathways were computed using the climbing-image nudged elastic band (CI-NEB) method~\cite{Henkelman2000} as implemented in ASE. Each NEB chain consisted of 11 images interpolated linearly between relaxed initial and final states. The chain was optimized using the BFGS algorithm until the maximum force perpendicular
to the band fell below 0.05~eV~\AA$^{-1}$. Transition-state
geometries and energies were extracted from the highest-energy
image of each converged chain.

\subsection*{Morse constraint on active site counts}

The covalent field $\Phi^{\mathcal{R}}_{\mathcal{G}}\vert_{\mathcal{M}}$, evaluated by a smooth energy oracle on a compact orientable manifold, is generically a Morse function. The Morse inequalities then impose a global constraint on its critical points:

\begin{equation}
    c_0 - c_1 + c_2 = \chi(\mathcal{M}),
\end{equation}

where $c_k$ counts critical points of index $k$: binding minima ($k=0$), migration saddles ($k=1$), and repulsive maxima ($k=2$), and $\chi(\mathcal{M})$ is the Euler characteristic of the manifold. For a nanoparticle with spherical topology $\chi(\mathcal{M}) = 2$, binding minima and saddle connections are not independent: the active site count is a global property of the field topology rather than a local geometric one. This constraint holds for the frozen probe scalar field; anisotropic and spectral extensions modify the topology of the domain but preserve the same organizing principle.

\subsection*{Linear scaling relation analysis}
 
Covalent field values for each probe family ($*\text{O}$/$*\text{OH}$; $*\text{N}$/$*\text{NH}$/$*\text{NH}_2$; $*\text{C}$/$*\text{CH}$/$*\text{CH}_2$/$*\text{CH}_3$) were evaluated at identical manifold vertices using frozen probes. Geometric equivalence across probe species within each family was enforced by fixing the anchor position and surface-normal orientation and varying only the molecular graph. Linear fits were performed by ordinary least squares. Deviations from the fitted line were computed per vertex and projected back onto the manifold surface for spatial visualization. For the HEO multi-channel descriptor, covalent field values were normalized to the $[0,1]$ range independently for each probe species before
assignment to the red ($*\text{O}$), green ($*\text{H}$), and blue ($*\text{CO}$) channels.

\subsection*{Point of Maximal Deviation (PMD)}

The PMD is distinct from the transition state: the transition state geometry depends on the thermodynamics of the specific initial and final states, whereas the PMD is the configuration of maximum higher-body-order residual along the reaction pathway, corrected for thermodynamic asymmetries. The change in molecular graph along the reaction coordinate identifies the forming bond directly, which defines the anchor of the PMD probe at its geometric midpoint, (consistent with a bidentate probe). The deviation from field additivity is largest when inter-probe interactions are strongest, and covalent interactions are inherently short-ranged and geometrically constrained.

The internal geometry of the PMD probe is therefore physically constrained for this reaction mechanism and can be approximated by a frozen conformation $\xi^*_0$, fixed across all manifold vertices. This approximation is justified not by the accuracy of the geometry itself but by a physical argument: what is transferred across manifold vertices is the requirement for simultaneous dual coordination, which is a property of the reaction class rather than of any specific site. The short-range character of covalent interactions ensures that this requirement imposes a geometrically narrow constraint that is largely insensitive to the identity of the anchoring metal. The energy scales confirm this directly: the distinction between sites in their capacity to stabilize simultaneous dual coordination is large relative to the orientational sensitivity of the frozen geometry, as shown by $\sigma_\alpha(\Phi^{\mathcal{M}}_{\text{PMD}})$ (\textbf{Figure~\ref{fig:bep_manifold}d}).

The gauge energy $E_0^{\text{PMD}}$ of the PMD probe is the interaction energy between probes $A$ and $B$ at the frozen geometry $\xi^*_0$ in the absence of the reference particle:

\begin{equation}
E_0^{\text{PMD}} = E\!\left(\mathcal{P}_A(\xi^*_0) \cup
\mathcal{P}_B(\xi^*_0)\right) - E_0^A - E_0^B.
\end{equation}

When $\Delta E_{\text{HBO}}^{\mathcal{M}}$ is used to identify active regions of the manifold, $E_0^{\text{PMD}}$ enters as a constant offset uniform across all vertices and does not affect the spatial distribution of inter-probe coupling.

Anisotropic field evaluation was performed by rotating the PMD probe about the surface normal in increments of $10^\circ$, yielding 36 energy evaluations 
per manifold vertex. The PMD field is:

\begin{equation}
\Phi^{\mathcal{R}}_{\text{PMD}}(\mathbf{r}; \alpha) =
E(\mathcal{R} \cup \mathcal{P}_{\text{PMD}}(\mathbf{r}, 
\alpha)) - E(\mathcal{R}) - E_0^{\text{PMD}},
\end{equation}

where $\alpha$ is rotation about the surface normal.

\section*{Data Availability}
The datasets used in this study are available via Zenodo \cite{zenodo}.

\section*{Code Availability}
The source code of the CFT framework is available on GitHub at: \url{https://github.com/schwallergroup/CFT}.

\section*{Acknowledgments}
E.F. and P.S. acknowledge support from the NCCR Catalysis (grant number 225147), a National Center of Competence in Research funded by the Swiss National Science Foundation.

\section*{Author contributions}

PS contributed to conceptualization, methodology, writing, editing, funding and project supervision. EF contributed to conceptualization, methodology, implementation, writing, visualization and assessment. 

\section*{Competing interests}
The authors declare no competing interests.


\clearpage
\nolinenumbers
\bibliographystyle{unsrt}
\bibliography{ref}

\clearpage
\setcounter{page}{1}
\renewcommand{\thefigure}{S\arabic{figure}}
\setcounter{figure}{0}

\begin{center}
    {\LARGE Supporting Information: On the Covalent Fields of Molecule–Surface
Interactions} \\[1.5em]  
    
    {\large Edvin Fako$^{1, 2, *}$, Philippe Schwaller$^{1, 2, *}$}\\[0.5em]
    
    {\small
    Corresponding authors: edvinfako@gmail.com (E. Fako), philippe.schwaller@epfl.ch (P. Schwaller)
    }
    
    {\small
    $^1$ Laboratory of Artificial Chemical Intelligence (LIAC), Institute of Chemical Sciences and Engineering, Ecole Polytechnique F\'{e}d\'{e}rale de Lausanne (EPFL), Lausanne, Switzerland

    $^2$ National Centre of Competence in Research (NCCR) Catalysis, Ecole Polytechnique F\'{e}d\'{e}rale de Lausanne (EPFL), Lausanne, Switzerland

    }
\end{center}

\vspace{2em}  



\newpage

\subsection*{Supplementary Note~1: Proto-CFT Analysis of the Ligand-Modulated Copper Electrocatalyst}

\begin{figure}[ht!]
\centering
\includegraphics[width=1.0\linewidth]{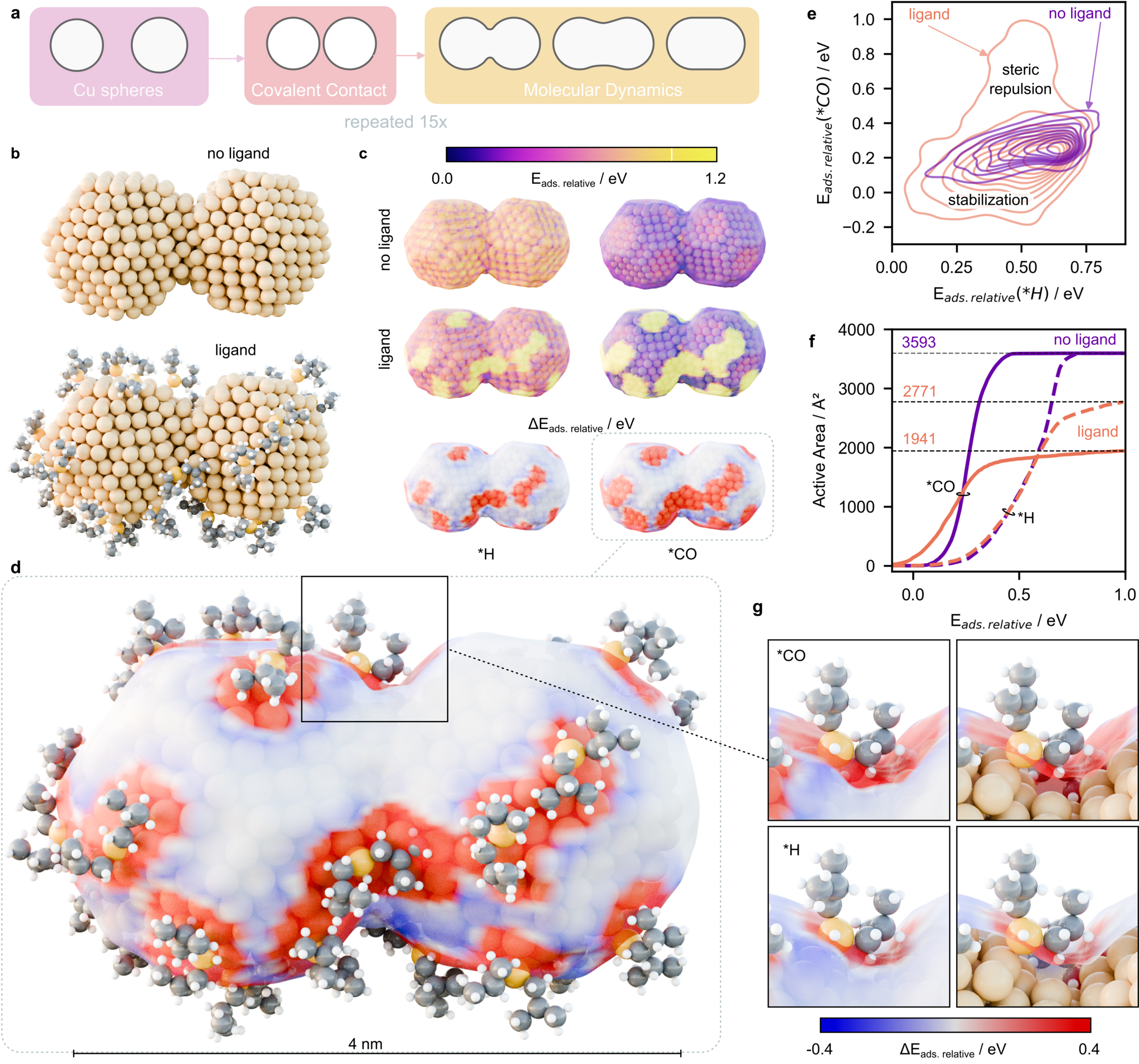}
\caption{\textbf{Proto-CFT analysis of ligand-modulated 
$*$CO and $*$H affinities on a grain-boundary forming Cu 
nanoparticle \cite{Leemans2026}.}
\textbf{a}, Workflow for constructing the Cu/Cu grain-boundary 
model: two Cu spheres are brought into contact, relaxed, and 
evolved by molecular dynamics to generate a representative 
ensemble of roughened surface motifs.
\textbf{b}, Atomistic snapshot of the grain-boundary interface 
with (orange) and without (purple) PR$_2$H ligand coverage at 
experimentally determined surface density ($\theta = 0.6$).
\textbf{c}, Manifold distributions of $*$CO and $*$H covalent 
field values with and without ligands; ligands broaden both 
distributions and shift the $*$CO distribution more severely 
toward weaker binding.
\textbf{d}, Delta affinity map 
$\Delta\Phi^{\mathcal{M}}_{*\text{CO}}$ projected onto the 
particle surface (blue: suppressed, red: enhanced); the spatial 
extent of ligand-induced $*$CO suppression is visible directly 
on the surface geometry.
\textbf{e}, Kernel density estimates of the manifold affinity 
distributions for $*$CO and $*$H with (orange) and without 
(purple) ligand coverage, showing asymmetric broadening.
\textbf{f}, Cumulative active area as a function of manifold 
affinity criterion for $*$CO and $*$H; $*$CO adsorption sites 
are suppressed more severely than $*$H sites, disfavoring 
C--C coupling during early catalyst activation.
\textbf{g}, Zoom into the local covalent field environment 
surrounding a single phosphine ligand for $*$H (left) and 
$*$CO (right); the larger spatial footprint of the $*$CO 
suppression region relative to $*$H is geometrically explicit 
in the field representation.
Adapted from Leemans et al., \textit{J. Am. Chem. Soc.} 
\textbf{148}, 13118--13127 (2026) under CC-BY~4.0.}
\label{SI_leemans}
\end{figure}

The atomistic simulations reported in Ref.~\cite{Leemans2026} constitute the first application of the manifold evaluation procedure and \texttt{AutoAdsorbate} probe placement that underlies CFT, predating its formal axiomatization. A shrinkwrap surface manifold was constructed over a grain-boundary forming Cu nanoparticle dimer (\textbf{Supplementary Figure~\ref{SI_leemans}a,b}), and the covalent fields of $*$CO and $*$H were evaluated at each manifold vertex in the presence and absence of secondary phosphine (PR$_2$H) ligands. The manifold distributions (\textbf{Supplementary Figure~\ref{SI_leemans}c}) and the delta affinity map $\Delta\Phi^{\mathcal{M}} = \Phi^{\mathcal{M}}_{\text{ligand}} - \Phi^{\mathcal{M}}_{\text{no ligand}}$ projected onto the particle surface (\textbf{Supplementary Figure~\ref{SI_leemans}d}) reveal that ligands introduce pronounced site heterogeneity: ligand-free sites acquire enhanced affinity for both intermediates while ligand-occupied regions are suppressed. The kernel density distributions (\textbf{Supplementary Figure~\ref{SI_leemans}e}) and cumulative active area analysis (\textbf{Supplementary Figure~\ref{SI_leemans}f}) show that this suppression is asymmetric: $*$CO adsorption sites are reduced more severely than $*$H sites, disfavoring the adjacent $*$CO intermediates required for C--C coupling and consistent with methane-dominated selectivity during early activation observed experimentally. The local origin of this asymmetry is directly readable from the covalent field maps in the immediate environment of a single ligand (\textbf{Supplementary Figure~\ref{SI_leemans}g}): the steric and electronic footprint of the phosphine suppresses the extended $*$CO binding region more than the compact $*$H binding region, a consequence of the different spatial demands of the two intermediates that the field representation makes geometrically explicit.

This analysis used the manifold machinery as a computational procedure that produced correct results. CFT provides the theoretical grounding that makes those results derived quantities of a formal theory rather than outputs of an ad hoc calculation: the asymmetric suppression of $*$CO and $*$H affinities is a direct consequence of the multi-channel field structure $(\Phi^{\mathcal{M}}_{*\text{H}}, \Phi^{\mathcal{M}}_{*\text{CO}})$, connected by the same 
representational choice to the Morse topology of active site counts, the manifold-level origin of linear scaling relations, and the algebraic identity underlying BEP correlations. Ref.~\citenum{Leemans2026} used proto-CFT to obtain the right answer; CFT explains why it was the right answer and what else 
follows from the same representational choice.

\newpage

\subsection*{Supplementary Figures}

\begin{figure*}[h!]
	\centering
    \includegraphics[width=1.0\linewidth]{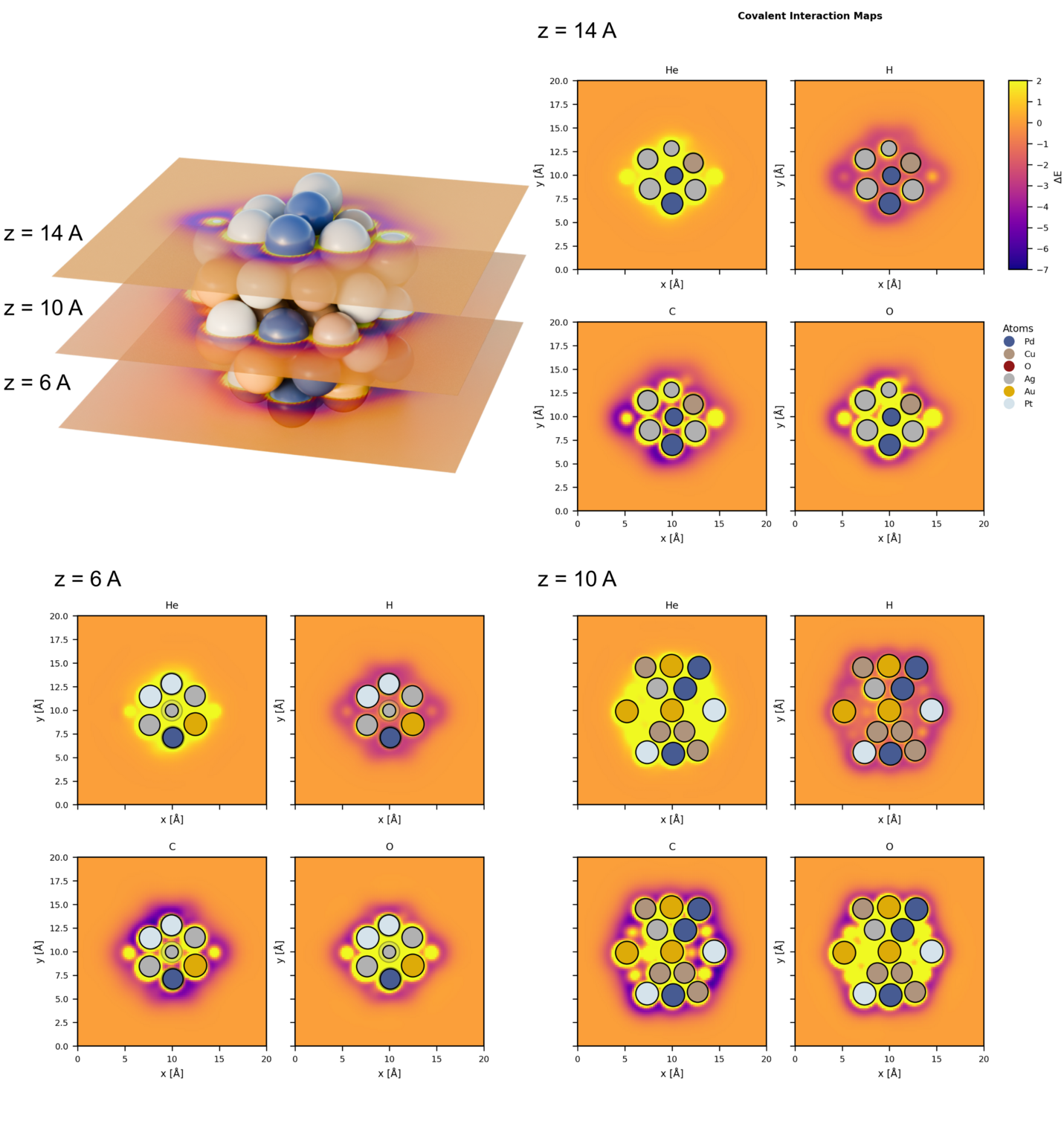}
\caption{\textbf{Covalent field slices for atomic probes on the
PtPdAuAgCu nanoparticle.}
Planar cross-sections of $\Phi^{\mathcal{R}}_{\mathcal{G}}$ 
evaluated at the same slice geometry as Fig.~1b for four 
atomic probes: $*\text{C}$, $*\text{O}$, $*\text{H}$, and 
$*\text{He}$ (yellow: repulsive, blue: strongly attractive). 
Carbon shows the deepest and most spatially confined field, 
consistent with strong, directional metal--carbon covalent 
bonding. The helium field is purely repulsive at all surface 
sites, confirming that attractive wells in the covalent probes 
reflect genuine covalent bonding rather than Pauli repulsion 
or dispersion.}
	\label{SI_slices}
\end{figure*}

\begin{figure*}[h!]
	\centering
    \includegraphics[width=1.0\linewidth]{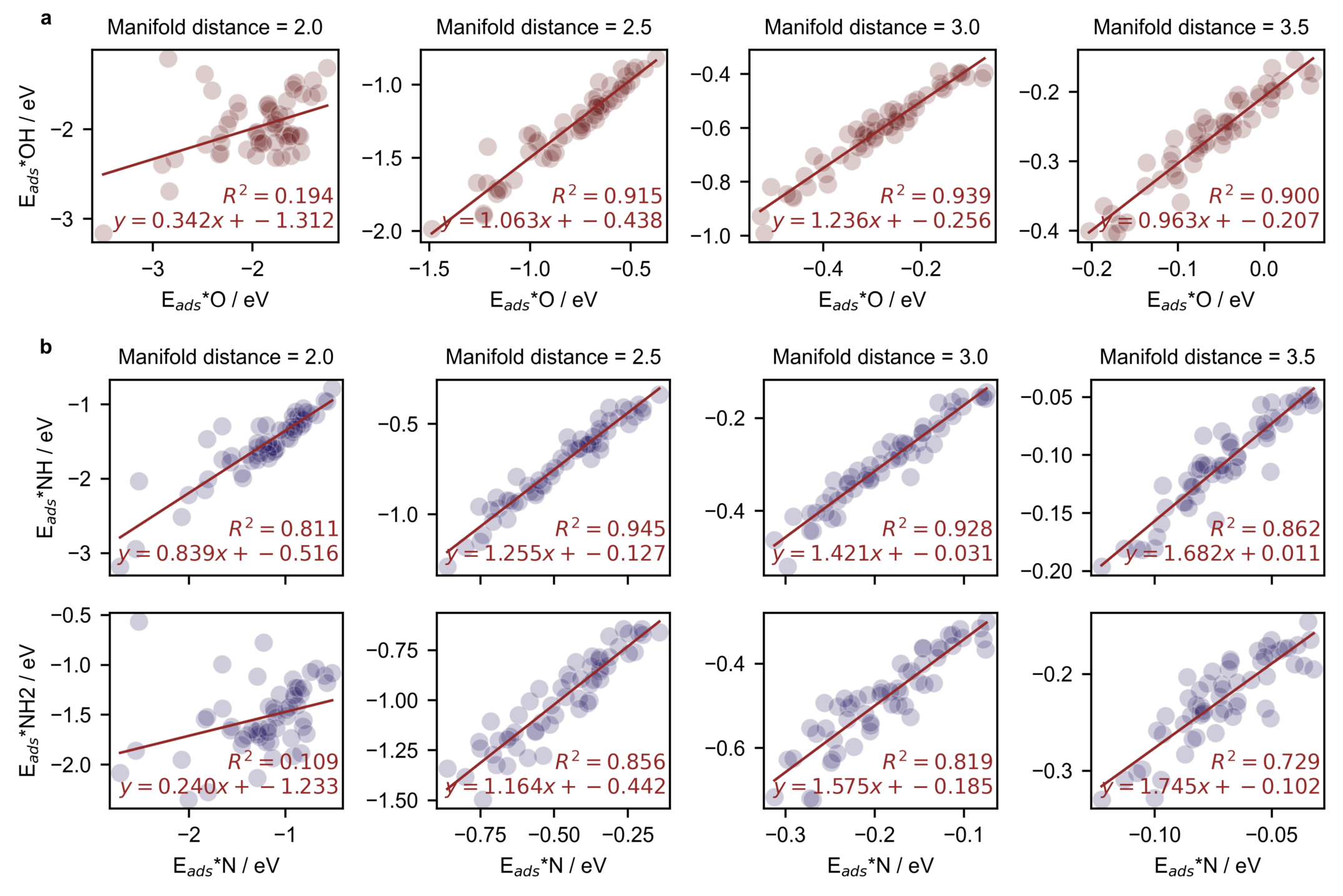}
\caption{\textbf{Linear scaling relations across manifold 
distances for the $*\text{O}$/$*\text{OH}$ and 
$*\text{N}$/$*\text{NH}$/$*\text{NH}_2$ probe families.}
Each panel corresponds to a static manifold evaluated at a 
fixed probe--surface distance (2.0, 2.5, 3.0, and 3.5~\AA). 
At 2.0~\AA\ the correlation is weaker, as probes at this 
distance sample a mixed regime of covalent and repulsive 
interactions. At 2.5~\AA\ and above, linear correlations 
are robust and stable, confirming that the scaling structure 
recovered by manifold sampling is insensitive to the exact 
choice of evaluation distance within the covalent interaction 
regime.}
	\label{SI_LSR1}
\end{figure*}

\begin{figure*}[h!]
	\centering
    \includegraphics[width=1.0\linewidth]{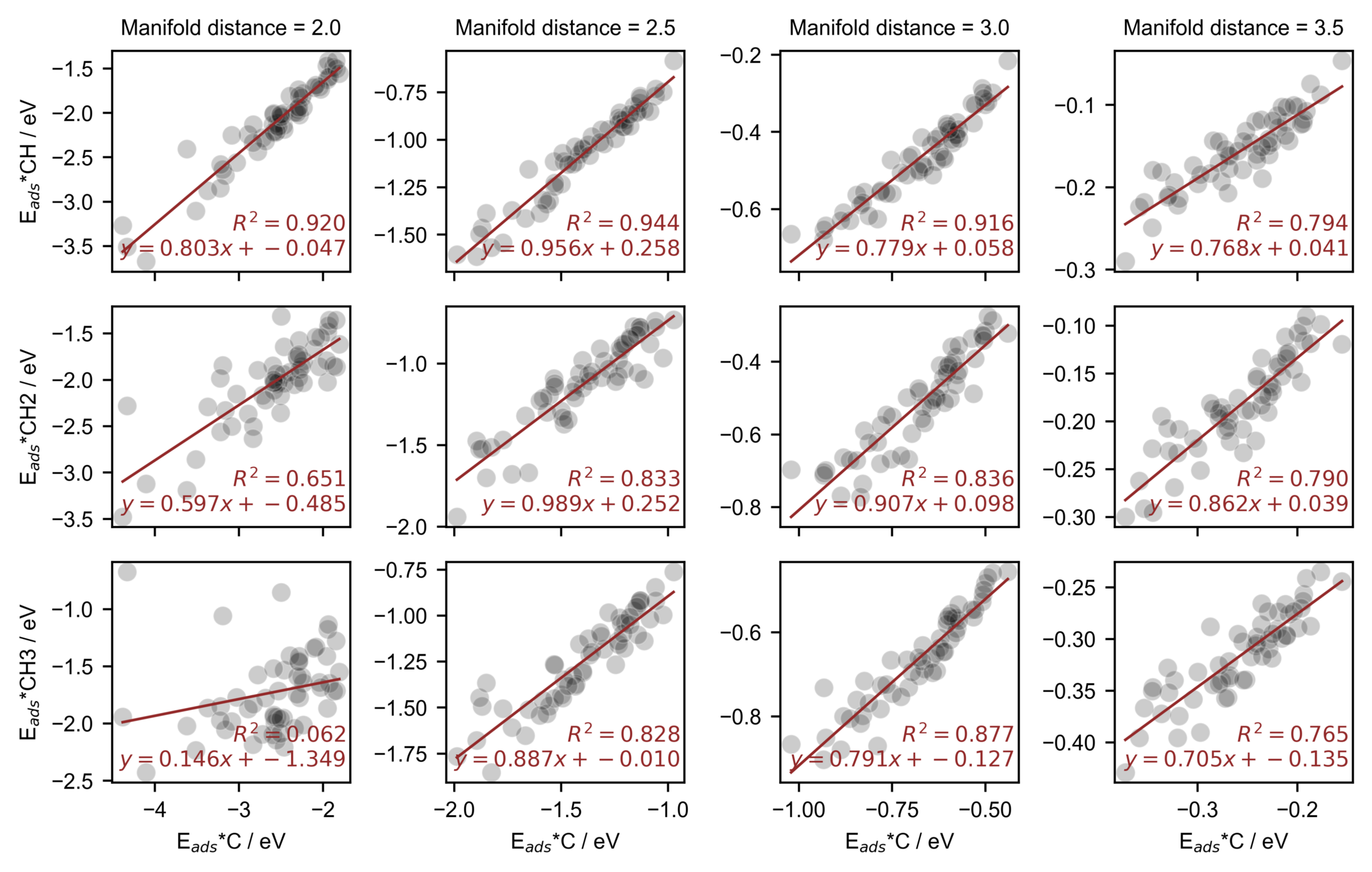}
\caption{\textbf{Linear scaling relations across manifold 
distances for the 
$*\text{C}$/$*\text{CH}$/$*\text{CH}_2$/$*\text{CH}_3$ 
probe family.}
As in Supplementary Fig.~\ref{SI_LSR1}, the 2.0~\AA\ 
manifold yields a weaker correlation while distances of 
2.5--3.5~\AA\ recover robust linear scaling trends. 
Consistent behavior across both probe families supports 
a fixed manifold evaluation distance in the range 
2.5--3.0~\AA\ for routine CFT analysis.}
	\label{SI_LSR2}
\end{figure*}

\begin{figure*}[h!]
	\centering
    \includegraphics[width=120mm]{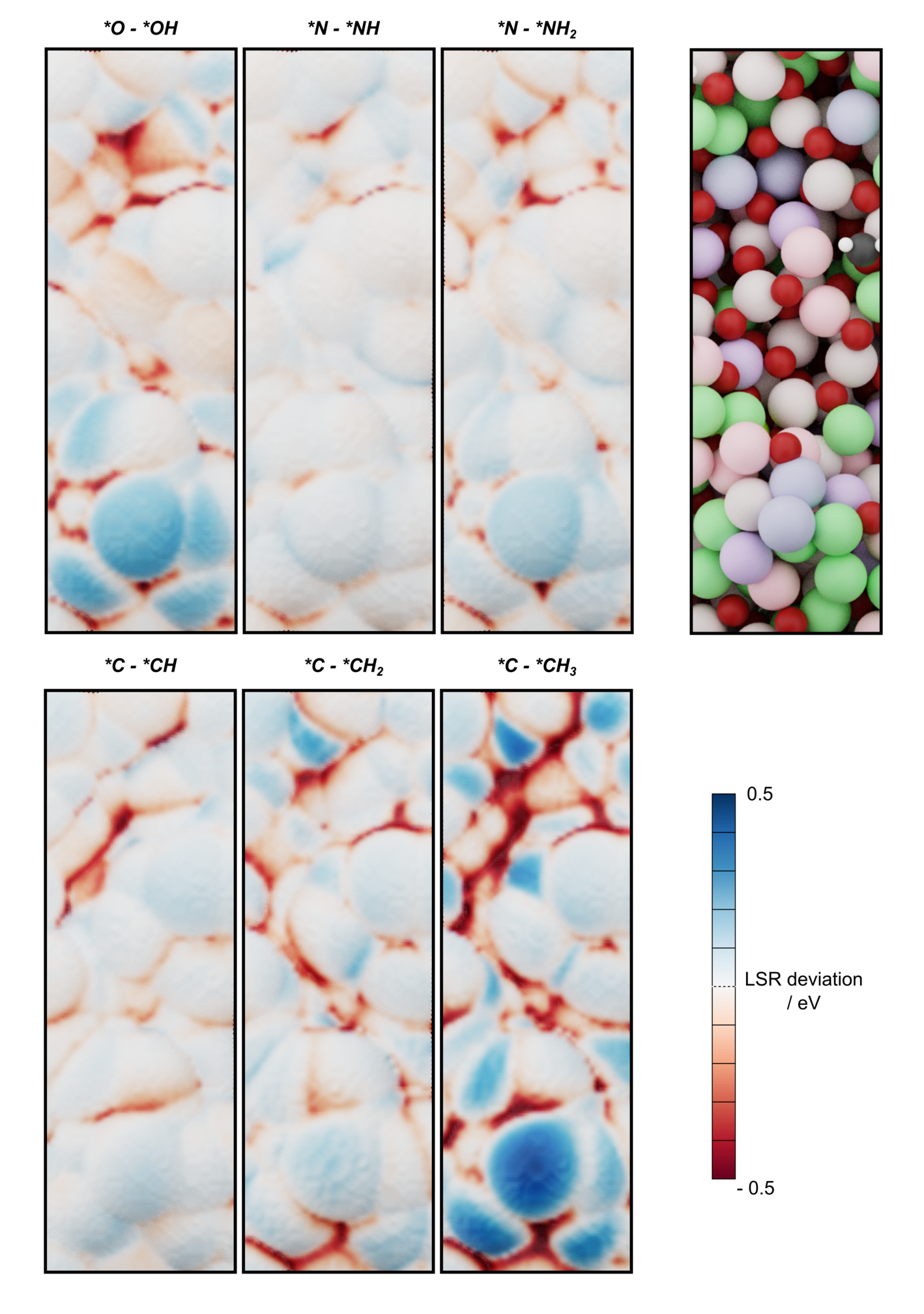}
\caption{\textbf{Spatially resolved linear scaling relation 
deviations on the high-entropy oxide surface.}
LSR deviation map for the partially reduced HEO slab 
(gray: on-trend, blue: weaker than predicted, red: stronger 
than predicted), constructed using the same procedure as 
Fig.~4d. The spatial distribution of deviations is markedly 
more heterogeneous than on the HEA nanoparticle, reflecting 
the chemical and structural diversity of the oxide surface. 
Fully oxidized patches appear blue; partially reduced 
metal-like sites appear red. The identical procedure applied 
to both systems confirms that manifold-resolved LSR analysis 
transfers directly to oxide surfaces without modification.}
	\label{SI_LSR_break_shuang}
\end{figure*}

\begin{figure*}[h!]
	\centering
    \includegraphics[width=120mm]{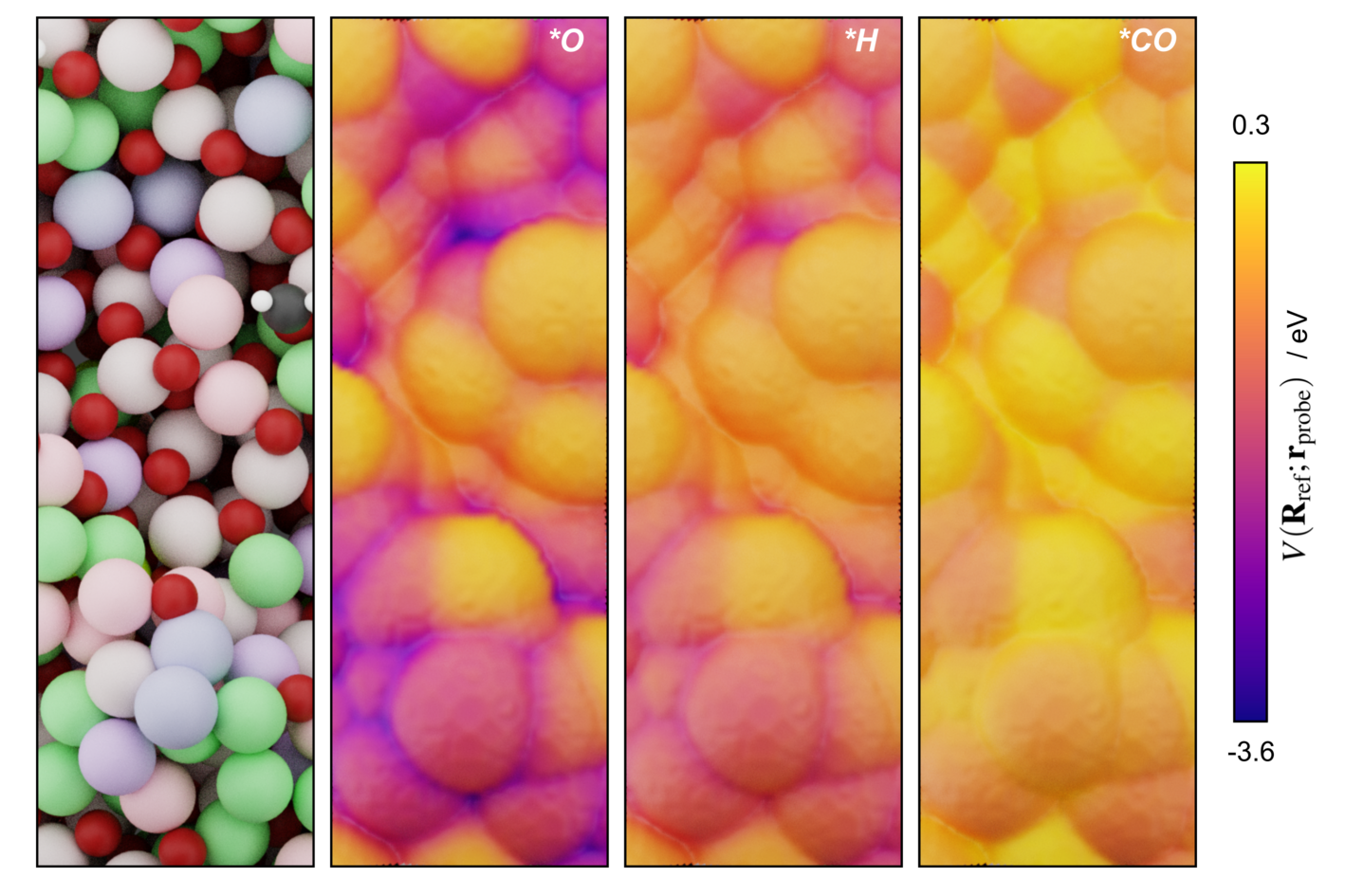}
\caption{\textbf{Individual channel maps of the multi-channel 
affinity descriptor on the partially reduced HEO surface.}
Covalent field distributions for $*\text{O}$ (red channel), 
$*\text{H}$ (green channel), and $*\text{CO}$ (blue channel) 
shown separately on a common energy scale, corresponding to 
the three components of the composite map in Fig.~4f. 
$*\text{H}$ and $*\text{O}$ affinities track each other 
closely across the surface while $*\text{CO}$ affinity 
varies independently, producing the spatial decoupling 
visible in the RGB composite.}
	\label{SI_bagger_2017}
\end{figure*}

\begin{figure*}[h!]
	\centering
    \includegraphics[width=1.0\linewidth]{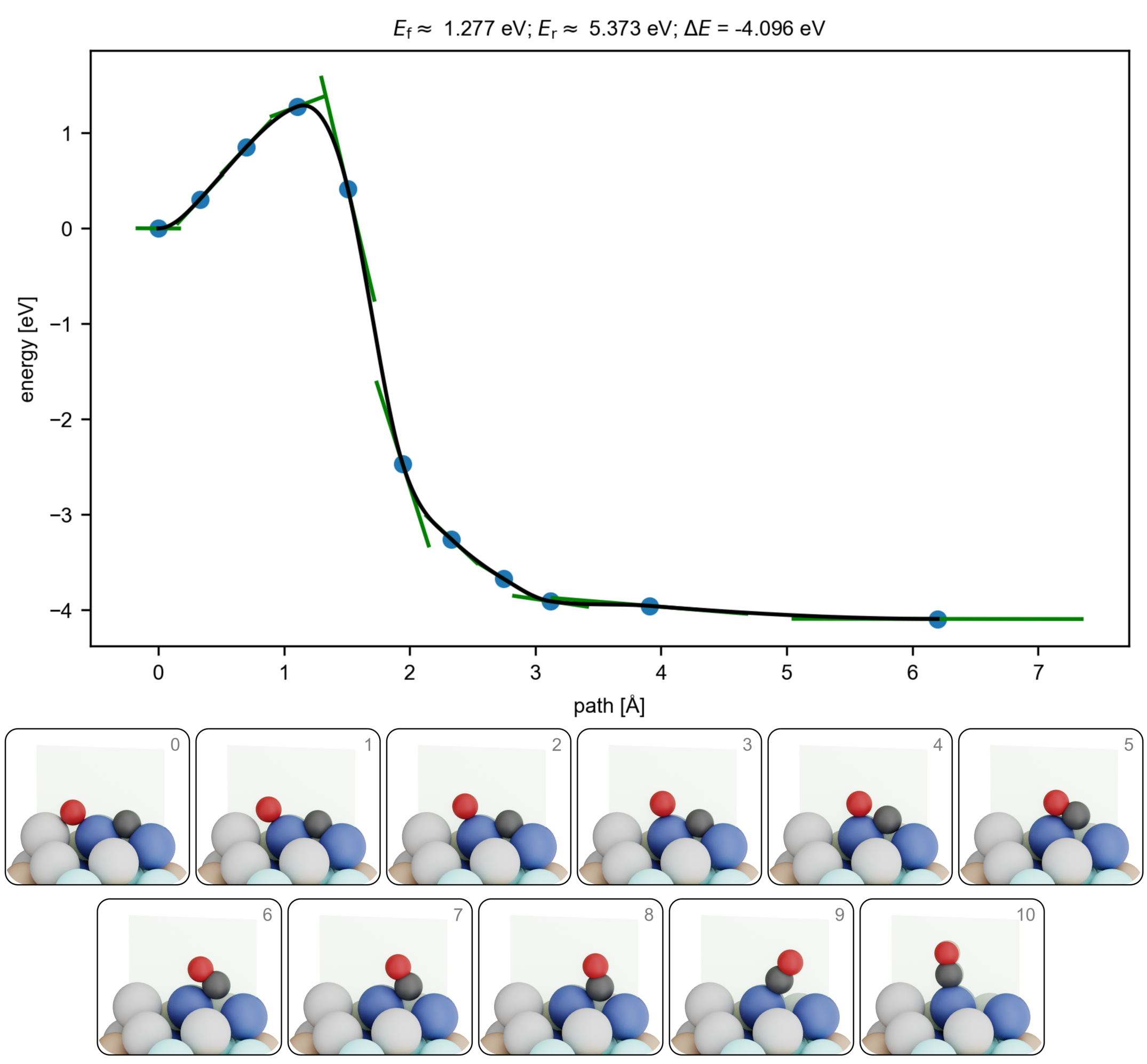}
\caption{\textbf{NEB pathway images for 
$*\text{C} + *\text{O} \rightarrow *\text{CO}$.}
Atomistic structures of all 11 NEB images along the 
minimum-energy pathway, with the energy profile shown above; 
images are colored by total energy relative to the initial 
state. The initial state (image~1) has C and O adsorbed on 
neighboring bridge sites of the same Pd atom. The transition 
state (image~3, TS) corresponds to O traversing the shared 
Pd top site. The final state (image~11) is top-adsorbed CO. 
The point of maximal deviation (image~5, PMD) marks the 
configuration where both C and O remain simultaneously 
within covalent range of the same surface atom; this point 
is path-independent and distinct from the transition state, 
which depends on the thermodynamics of the specific initial 
and final states.}
	\label{SI_neb}
\end{figure*}

\begin{figure*}[h!]
	\centering
    \includegraphics[width=1.0\linewidth]{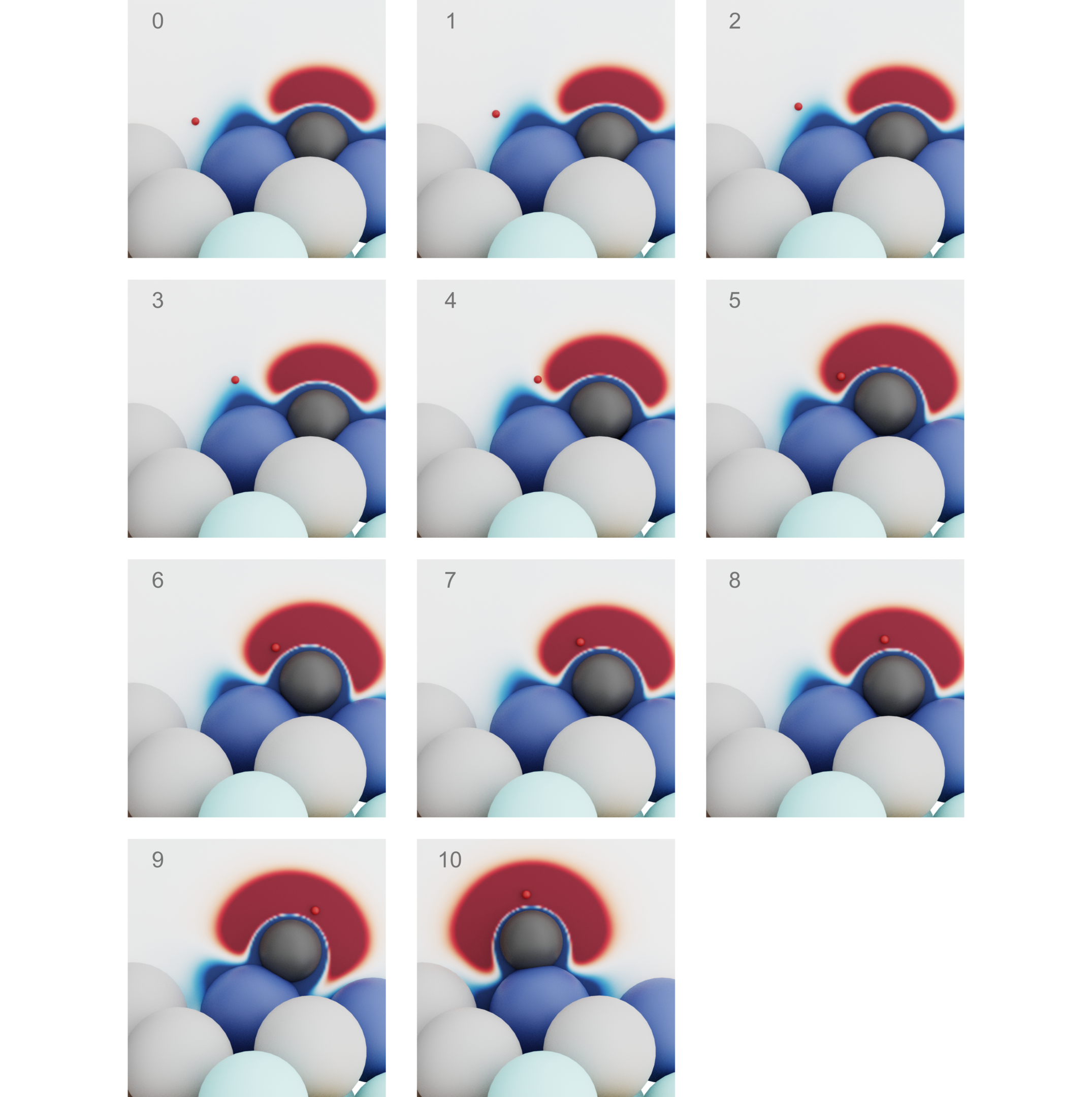}
\caption{\textbf{Perturbation of the $*\text{O}$ covalent 
field by the presence of $*\text{C}$ along the NEB pathway.}
Each panel shows the difference field 
$\Delta\Phi_{*\text{O}} = 
\Phi^{\mathcal{R}_{*\text{C}}}_{*\text{O}} - 
\Phi^{\mathcal{R}}_{*\text{O}}$ on a planar slice through 
the nanoparticle at each NEB image (blue: C destabilizes 
$*\text{O}$; red: C enhances $*\text{O}$ stability via 
incipient C--O bond formation); current O position marked 
in red. In the initial state (image~1) O sits in a white 
region, confirming negligible inter-probe coupling at large 
separation. A persistent deep-red attractive region is 
present in all images, indicating that C--O bond formation 
is energetically favorable across a wide range of O approach 
geometries. The transition state lies within a repulsive 
region, revealing that the conventional barrier originates 
in the limited capacity of the Pd atom to bind $*\text{C}$ 
and $*\text{O}$ simultaneously rather than in the intrinsic 
difficulty of C--O bond formation.}
	\label{SI_co_field}
\end{figure*}


\begin{figure}[h!]
    \centering
    \includegraphics[width=\textwidth]{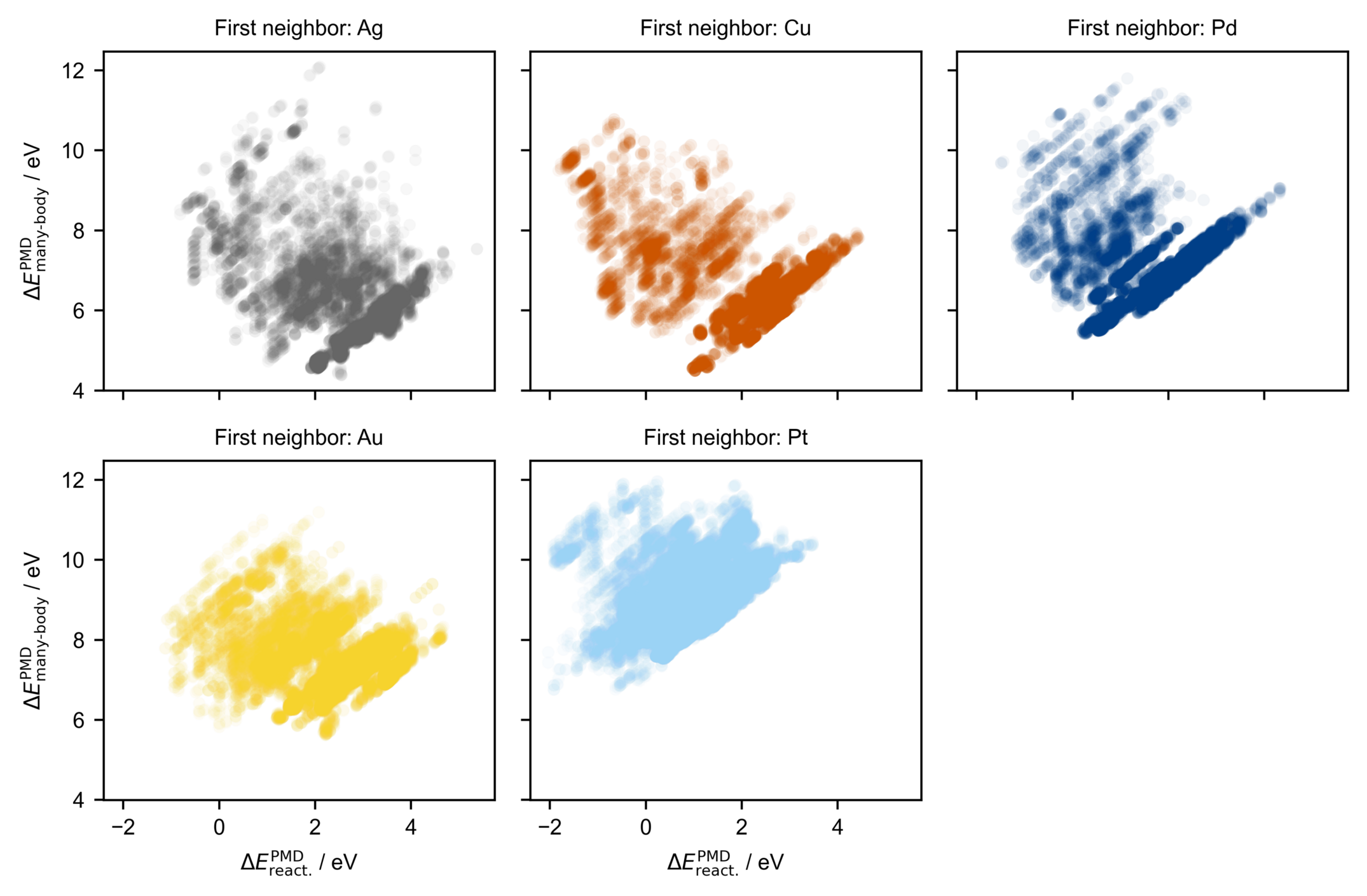}
\caption{\textbf{Apparent BEP correlations across the HEA 
nanoparticle, segmented by anchoring metal.}
Thermodynamics-corrected barrier 
($\Delta E_{\text{HBO}}^{\mathcal{M}} - E_{\text{react.}}^{\text{IS}}$) 
versus reaction energy for all $\sim$120,000 candidate 
pathways, colored by anchoring metal. Parallel families 
with unit slope emerge for each metal, consistent with 
the algebraic identity derived in Methods: varying the 
initial state shifts both axes equally, generating parallel 
lines whose vertical offsets reflect the intrinsic barrier 
capacity of each PMD basin.}
\label{fig:bep_reaction_barrier}
\end{figure}

\begin{figure}[h!]
    \centering
    \includegraphics[width=\textwidth]{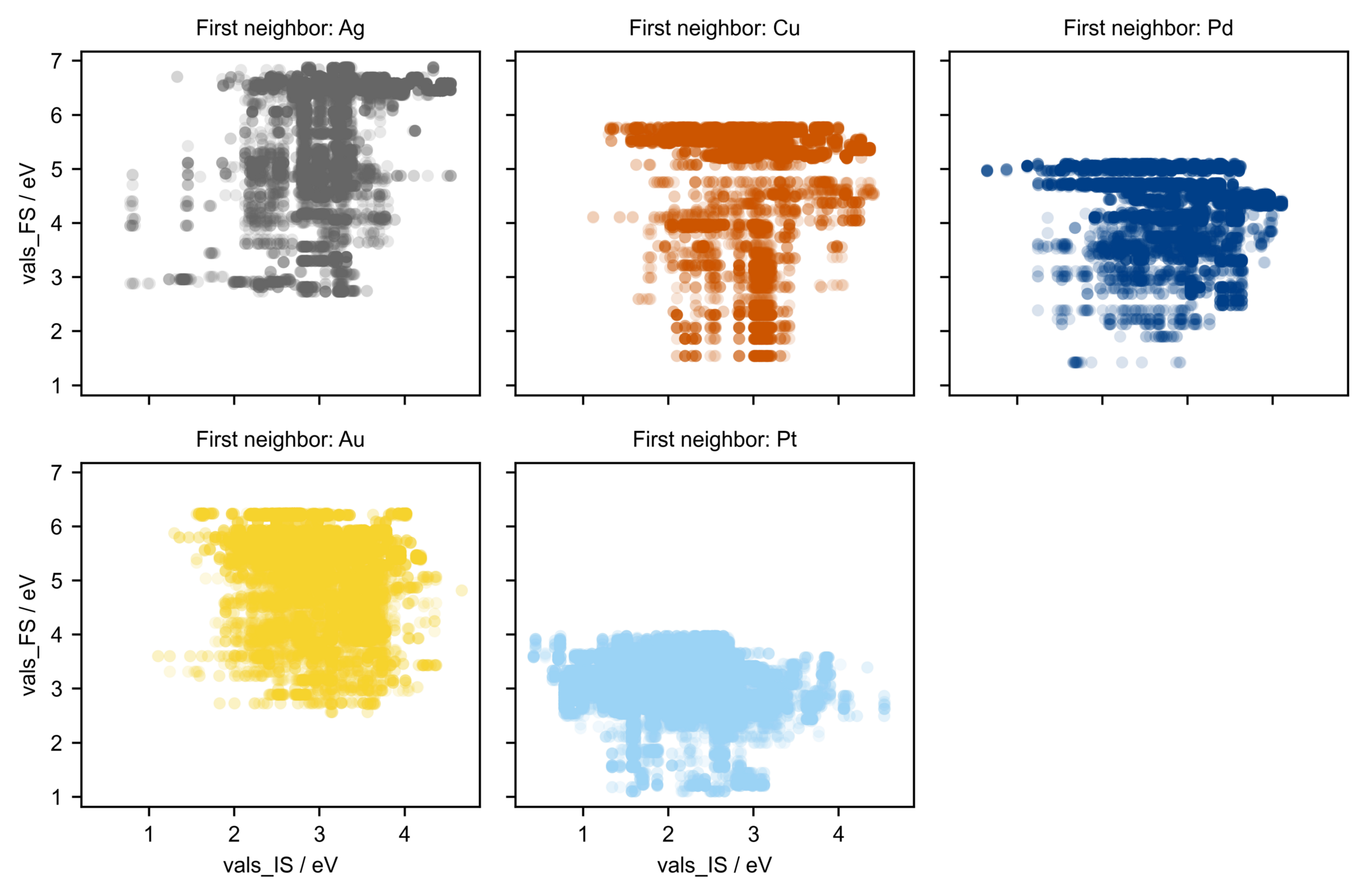} 
\caption{\textbf{Initial-state versus final-state covalent 
field energies across the HEA nanoparticle.}
Absence of correlation between $*\text{C}$/$*\text{O}$ 
initial-state affinities and $*\text{CO}$ final-state 
affinities confirms that initial-state configurations are 
distributed independently of the PMD basin they access, 
validating the decoupling assumed in the BEP decomposition.}
\label{fig:is_vs_fs}
\end{figure}

\begin{figure}[h!]
    \centering
    \includegraphics[width=\textwidth]{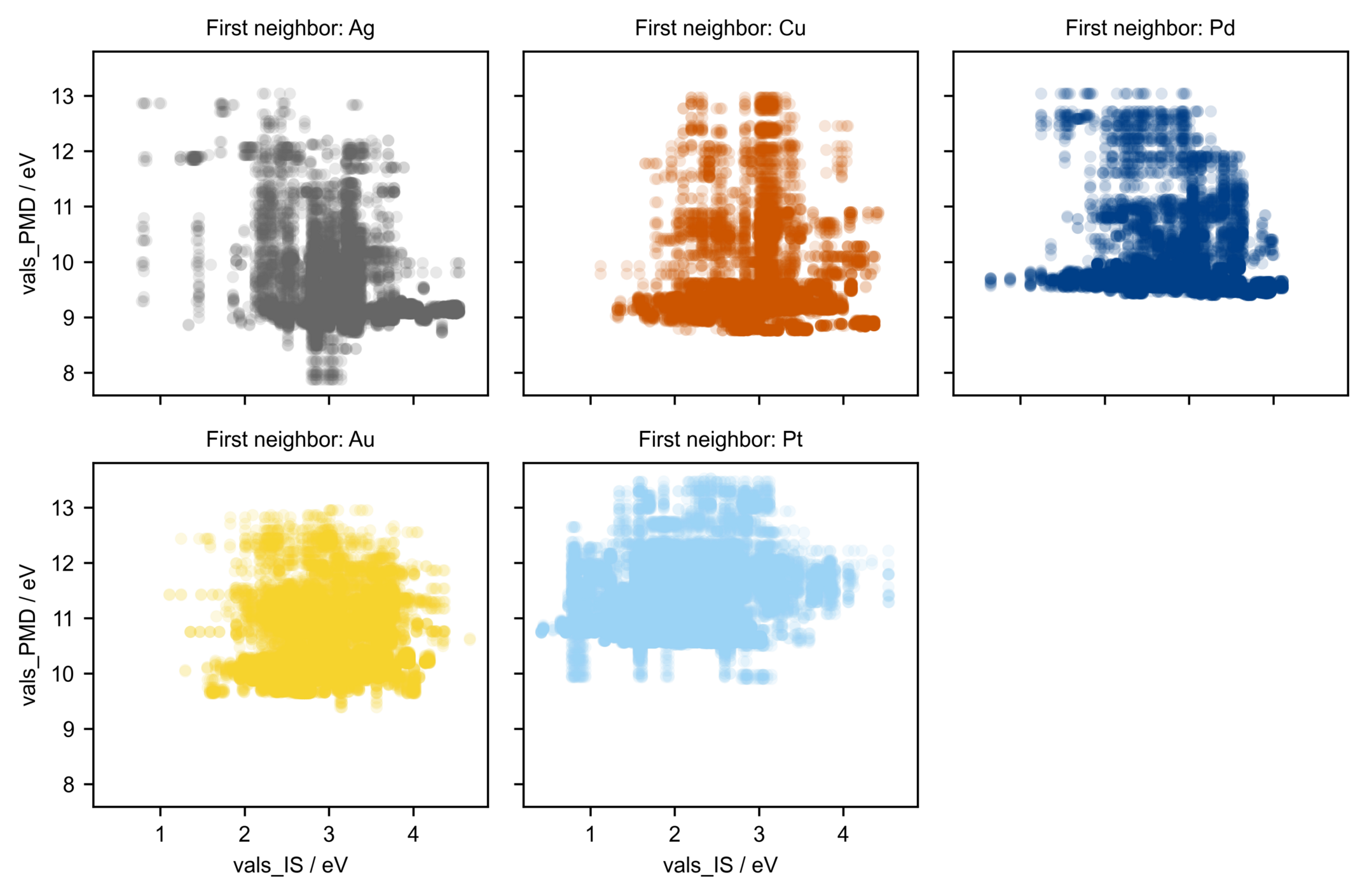} 
\caption{\textbf{Initial-state covalent field energy versus 
$\Phi _{\text{HBO}}^{\text{PMD}}$ across the HEA 
nanoparticle.}
Absence of correlation confirms that the thermodynamic 
stability of the initial $*\text{C}$ and $*\text{O}$ 
configurations is independent of the intrinsic barrier 
capacity of the PMD basin, consistent with the 
path-independence of $\Delta E_{\text{HBO}}^{\mathcal{M}}$ 
established in Methods.}
\label{fig:is_vs_pmd}
\end{figure}

\begin{figure}[h!]
    \centering
    \includegraphics[width=\textwidth]{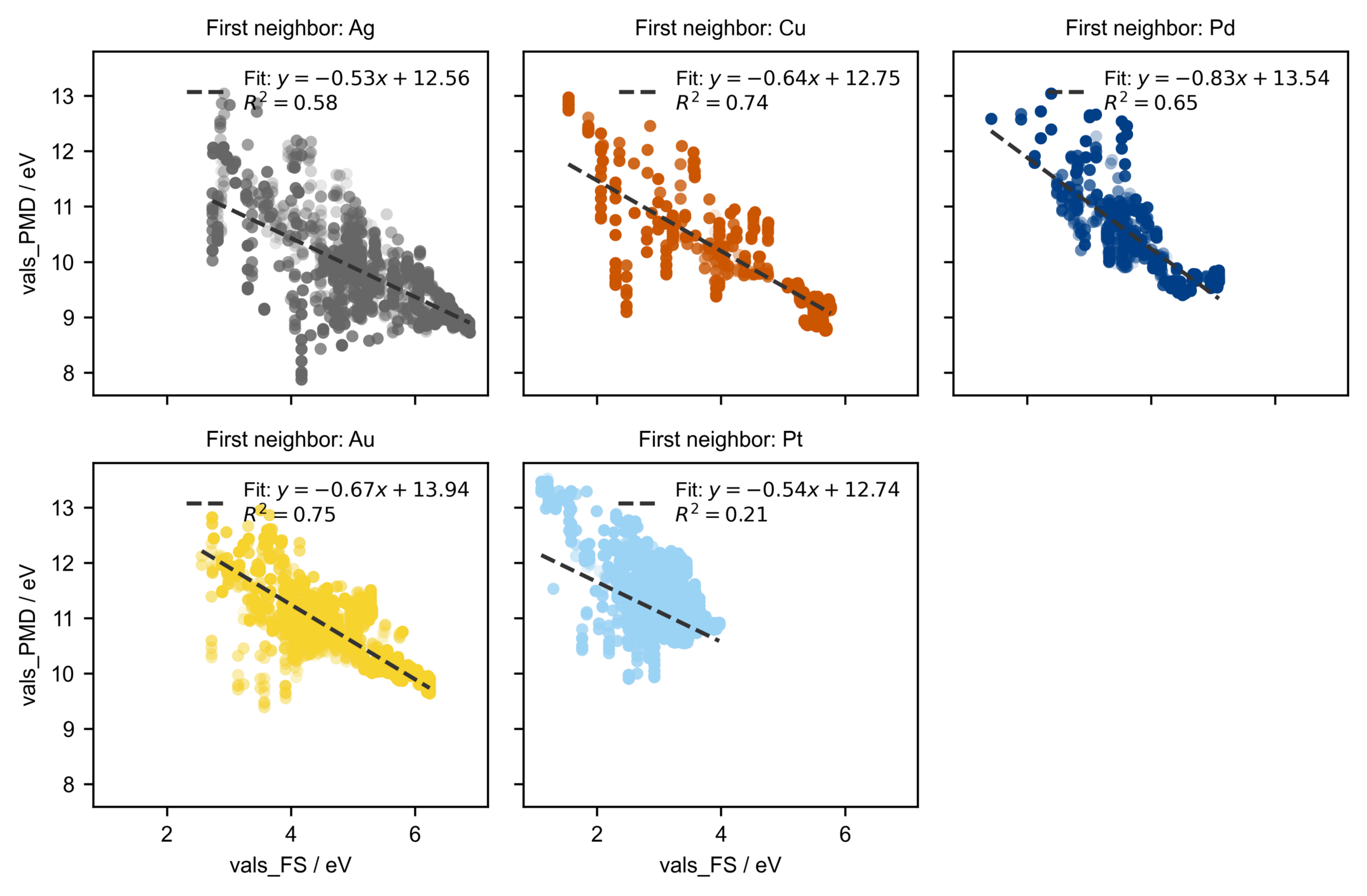} 
\caption{\textbf{Final-state $*\text{CO}$ affinity versus 
$\Delta E_{\text{HBO}}^{\mathcal{M}}$ across the HEA 
nanoparticle.}
Inverse correlation between final-state $*\text{CO}$ 
binding and intrinsic barrier capacity reflects a geometric 
incompatibility between the simultaneous dual coordination 
demanded by the PMD geometry and the top-site preference 
of adsorbed CO: sites most effective at stabilizing the 
strained PMD configuration bind the product more weakly. 
This anti-correlation is a manifold-level property 
invisible to site-averaged analysis.}
\label{fig:fs_vs_pmd}
\end{figure}

\clearpage

\end{document}